\documentclass[sigconf]{_acm/acmart}

\usepackage{bigstrut}
\usepackage{booktabs}
\usepackage{multirow}
\usepackage{microtype}
\usepackage{graphicx}
\usepackage{bbding}
\usepackage{overpic}
\usepackage{balance}
\usepackage{amsthm}
\usepackage{array}
\usepackage{clrscode}
\usepackage{multicol}
\usepackage{float}
\usepackage{newfloat}
\usepackage{color}
\usepackage{xcolor}
\usepackage{amsopn}
\usepackage{mathrsfs}
\usepackage{mathtools}
\usepackage{amsmath}
\usepackage{arydshln}
\usepackage{blkarray}
\usepackage{enumerate}
\usepackage{courier}
\usepackage{rotating}
\usepackage{bm}
\usepackage{wrapfig}
\usepackage{ragged2e}
\usepackage[misc]{ifsym}
\usepackage{threeparttable}
\usepackage{makecell}
\usepackage{adjustbox} 
\usepackage{enumitem}
\usepackage{hyperref}
\usepackage{url}
\usepackage{xspace}
\usepackage{comment}
\usepackage{changepage}
\usepackage{algorithm}
\usepackage{algpseudocode} 
\usepackage{algorithmicx} 
\usepackage[ruled, vlined, nofillcomment, linesnumbered, algo2e]{algorithm2e}
\usepackage[normalem]{ulem}
\usepackage{caption}
\usepackage{subcaption}
\newcommand{\ul}{\underline}
\newcommand{\ie}{\emph{i.e.,}\xspace}
\newcommand{\eg}{\emph{e.g.,}\xspace}

\newcommand{\paratitle}[1]{\vspace{1.5ex}\noindent\textbf{#1}}

\newcommand{\ignore}[1]{}

\AtBeginDocument{%
  \providecommand\BibTeX{{%
    \normalfont B\kern-0.5em{\scshape i\kern-0.25em b}\kern-0.8em\TeX}}}

\copyrightyear{2025} 
\acmYear{2025} 
\setcopyright{acmlicensed}\acmConference[SIGIR '25]{Proceedings of the 48th International ACM SIGIR Conference on Research and Development in Information Retrieval}{July 13--18, 2025}{Padua, Italy}
\acmBooktitle{Proceedings of the 48th International ACM SIGIR Conference on Research and Development in Information Retrieval (SIGIR '25), July 13--18, 2025, Padua, Italy}
\acmDOI{10.1145/3726302.3729989}
\acmISBN{979-8-4007-1592-1/2025/07}

\begin{document}

\title{Generative Recommender with End-to-End Learnable \\ Item Tokenization}

\author{Enze Liu$^*$}
\orcid{0009-0007-8344-4780}
\affiliation{%
    \department{Gaoling School of Artificial Intelligence}
    \institution{Renmin University of China}
    \city{Beijing}
    \country{China}
}
\email{enzeliu@ruc.edu.cn}

\author{Bowen Zheng$^*$}
\orcid{0009-0002-3010-7899}
\affiliation{%
    \department{Gaoling School of Artificial Intelligence}
    \institution{Renmin University of China}
    \city{Beijing}
    \country{China}
}
\email{bwzheng0324@ruc.edu.cn}

\author{Cheng Ling}
\affiliation{%
    \institution{Kuaishou Technology}
    \city{Beijing}
    \country{China}
}
\email{lingcheng@kuaishou.com}

\author{Lantao Hu}
\affiliation{%
    \institution{Kuaishou Technology}
    \city{Beijing}
    \country{China}
}
\email{hulantao@gmail.com}

\author{Han Li}
\affiliation{%
    \institution{Kuaishou Technology}
    \city{Beijing}
    \country{China}
}
\email{lihan08@kuaishou.com}

\author{Wayne Xin Zhao
\textsuperscript{\Letter}
}

\orcid{0000-0002-8333-6196}
\affiliation{
    \department{Gaoling School of Artificial Intelligence}
    \institution{Renmin University of China}
    \city{Beijing}
    \country{China}
}
\email{batmanfly@gmail.com}

\thanks{$^*$ Equal contribution.}
\thanks{\Letter \ Corresponding author.}

\renewcommand{\shortauthors}{Enze Liu, et al.}
\renewcommand{\shorttitle}{Generative Recommender with End-to-End Learnable Item Tokenization}

\begin{abstract}
    Generative recommender systems have gained increasing attention as an innovative approach that directly generates item identifiers for recommendation tasks. Despite their potential, a major challenge is the effective construction of item identifiers that align well with recommender systems. Current approaches often treat item tokenization and generative recommendation training as separate processes, which can lead to suboptimal performance. To overcome this issue, we introduce \textbf{ETEGRec}, a novel \underline{E}nd-\underline{T}o-\underline{E}nd \underline{G}enerative \underline{Rec}ommender that unifies item tokenization and generative recommendation into a cohesive framework. Built on a dual encoder-decoder architecture, ETEGRec consists of an item tokenizer and a generative recommender. To enable synergistic interaction between these components, we propose a recommendation-oriented alignment strategy, which includes two key optimization objectives: sequence-item alignment and preference-semantic alignment. These objectives tightly couple the learning processes of the item tokenizer and the generative recommender, fostering mutual enhancement. Additionally, we develop an alternating optimization technique to ensure stable and efficient end-to-end training of the entire framework. Extensive experiments demonstrate the superior performance of our approach compared to traditional sequential recommendation models and existing generative recommendation baselines.
    Our code is available at \href{https://github.com/RUCAIBox/ETEGRec}{{https://github.com/RUCAIBox/ETEGRec}}.
\end{abstract}



\maketitle

\section{Introduction}
\label{sec:introduction}

In recommender systems, it is essential to model the sequential patterns of user behaviors, so as to effectively predict the future interactions for target users. Such task setting is typically formulated as \emph{sequential recommendation}~\cite{DBLP:journals/corr/HidasiKBT15,sasrec,DBLP:conf/www/RendleFS10,DBLP:conf/wsdm/TangW18}, which has attracted increasing research attention. 
Traditional sequential recommenders~\cite{sasrec, bert4rec, DBLP:journals/corr/HidasiKBT15} are often developed based on sequence models (\eg RNN, CNN, and Transformer) and make the predictions in a \emph{discriminative} way, \ie 
evaluating the similarity between the observed past sequence and candidate items, then selecting the most similar item(s) for recommendation.

Recently, drawing from the promising potential of generative language models~\cite{DBLP:conf/nips/0001YCWZRCYRR23,autoindexer}, several studies have emerged to apply the generative paradigm in recommender systems~\cite{tiger,eager,idgenrec,letter}.
Different from discriminative methods, the generative approach formulates the sequential recommendation task as a sequence-to-sequence problem and autoregressively generates the identifiers of target items.
Specifically, it generally involves two main aspects, namely \emph{item tokenization} and \emph{autoregressive generation}, for developing the entire recommendation framework. 
For item tokenization, it basically refers to assigning a list of meaningful IDs for indexing or representing an item. 
Existing efforts include parameter-free methods based on the co-occurrence matrix~\cite{gptrec,DBLP:conf/sigir-ap/HuaXGZ23}, text feature identifier~\cite{gpt4rec,DBLP:conf/recsys/Palma23,idgenrec}, hierarchical clustering~\cite{seater,eager}, and multi-level vector quantization (VQ)~\cite{tiger,tokenrec,letter,mmgrec}. 
Furthermore, several recent studies attempt to improve the quality of item identifiers by introducing collaborative signals~\cite{letter}, diversified regularization~\cite{letter}, or multi-behavior information~\cite{eager,mbgr}.
For autoregressive generation, the encoder-decoder architecture (\eg T5~\cite{t5}) is the most widely used backbone due to its excellent capabilities in sequence modeling and generation.
In addition, there are also some studies that aim to improve performance by adjusting the backbone architecture~\cite{seater,eager} or the learning objectives~\cite{seater}.

Despite these advancements, existing approaches typically consider item tokenization as a {pre-processing step} for subsequent generative recommendation.
This results in a \emph{complete decoupling} of the item tokenization and autoregressive generation during model optimization, which likely hinders the potential of generative recommendation due to two major reasons.
Firstly, the item tokenizer is essentially unaware of the optimization objectives for recommendation or simply not the best match for the recommender.
Secondly, the generative recommender cannot deeply fuse or further refine the prior knowledge implicitly encoded in item representations from the item tokenizer.
In light of these concerns, we aim to develop an end-to-end generative recommendation framework that seamlessly integrates item tokenization and autoregressive generation.
To realize this seamless integration of tokenization and generation, we highlight two primary challenges:
(1) How to integrate the item tokenizer and generative recommender into a unified recommendation framework; 
(2) How to achieve mutual enhancement between the item tokenizer and generative recommender for end-to-end optimization.

To this end, in this paper, we propose \textbf{ETEGRec}, an \underline{E}nd-\underline{T}o-\underline{E}nd \underline{G}enerative \underline{Rec}ommender that seamlessly integrates item tokenization and autoregressive generation.
Our framework adopts a dual encoder-decoder architecture, where the item tokenizer adopts Residual Quantization Variational Autoencoder (RQ-VAE), and the generative recommender is a Transformer model similar to T5.
The key novelty of our approach lies in that the item tokenizer can be jointly optimized with the generative recommender, which significantly differs from prior studies that use heuristic or pre-learned item tokenizers. 
In order to achieve mutual enhancement between these two components, we design two recommendation-oriented alignment strategies, which include \emph{sequence-item alignment} and \emph{preference-semantic alignment}.
Specifically, sequence-item alignment requires that the quantized token distributions from the encoder's sequential states and the collaborative embedding of the target item should be similar, 
and preference-semantic alignment employs contrastive learning to align the user preference captured by the Transformer decoder with the target item semantics reconstructed by RQ-VAE. 
Generally, our objective is to seamlessly integrate the tokenizer with the recommender through a well-crafted alignment approach, thereby fostering the mutual enhancement between the two components. 
Finally, to ensure stable and effective end-to-end learning, we further devise an alternating optimization method for joint training.

In summary, our main contributions are as follows:

$\bullet$ We propose a novel end-to-end generative recommender that achieves mutual enhancement and joint optimization of item tokenization and autoregressive generation.

$\bullet$ We design a recommendation-oriented alignment approach that mutually enhances the item tokenizer and the generative recommender through sequence-item alignment and preference-semantic alignment.

$\bullet$ We conduct extensive experiments on three recommendation benchmarks, demonstrating the superiority of our proposed framework compared with both traditional sequential recommendation models and generative recommendation baselines.

\section{Related Work}
\label{sec:related}

In this paper, we review the related work in two major aspects. 

\paratitle{Sequential Recommendation.}
Sequential recommendation aims to predict the next item a user may interact with based on the user's historical behavior sequences. 
Early studies~\cite{DBLP:conf/www/RendleFS10} primarily adhere to the Markov Chain assumption and focus on estimating the transition matrix. 
With the development of neural networks, various model architectures, such as Recurrent Neural Networks (RNN)~\cite{DBLP:journals/corr/HidasiKBT15, DBLP:conf/recsys/TanXL16}, Convolutional Neural Networks (CNN)~\cite{DBLP:conf/wsdm/TangW18} and Graph Neural Networks (GNN)~\cite{DBLP:conf/sigir/ChangGZHNSJ021, DBLP:conf/aaai/WuT0WXT19}, are applied for sequential recommendation. 
Recently, Transformer~\cite{transformer}-based recommendation models~\cite{sasrec, bert4rec, DBLP:journals/tkde/HaoZZLSXLZ23, s3rec} have achieved great success for effective sequential user modeling. SASRec~\cite{sasrec} utilizes Transformer decoder with unidirectional self-attention to capture user preference. 
BERT4Rec~\cite{bert4rec} proposes to encode the sequence by bidirectional attention and adopts the mask prediction task for training. 
S$^3$Rec~\cite{s3rec} explores using the intrinsic data correlation as supervised signals to pre-train the sequential model for better user and item representations. 
Furthermore, several works~\cite{fdsa, DBLP:conf/sigir/XieZK22} exploit the abundant textual features of users and items to enrich the user and item representations. In this work, we focus on exploring the generative paradigm for sequential recommendation.

\paratitle{Generative Recommendation}.  
Nowadays, generative recommendation has emerged as a next-generation paradigm for recommendation systems. 
In such a generative paradigm, the item sequence is tokenized into a token sequence and then fed into generative models to predict the tokens of the target item autoregressively.
Generally, the generative paradigm can be considered as two main processes, \ie item tokenization and generative recommendation.
Existing approaches for item tokenization can be broadly categorized into parameter-free methods~\cite{gptrec, DBLP:conf/sigir-ap/HuaXGZ23, DBLP:journals/corr/abs-2311-02089, idgenrec,seater,eager}, and deep learning methods based on multi-level vector quantization (VQ)~\cite{tiger, DBLP:journals/corr/abs-2403-18480, tokenrec, letter}.
For parameter-free methods, some studies, \eg CID~\cite{DBLP:conf/sigir-ap/HuaXGZ23} and GPTRec~\cite{gptrec}, apply matrix factorization to the co-occurrence matrix to derive item identifiers. 
Other works like SEATER~\cite{seater} and EAGER~\cite{eager} employ clustering of item embeddings to construct identifiers hierarchically. In addition, there are also some attempts~\cite{gpt4rec, DBLP:conf/recsys/Palma23, DBLP:conf/recsys/HarteZLKJF23, DBLP:journals/corr/abs-2311-02089, idgenrec} to use the textual metadata attached to items, \eg titles and descriptions, as identifiers. While these non-parametric methods are highly efficient, they often suffer from length bias and fail to capture deeper collaborative relationships among items. Deep learning methods based on multi-level VQ instead develop more expressive item identifiers with equal length via the Deep Neural Networks (DNN). For instance, TIGER~\cite{tiger} uses RQ-VAE to learn the codebooks. LETTER~\cite{letter} proposes to align quantized embeddings in RQ-VAE with collaborative embeddings to leverage both collaborative and semantic information.

Reviewing the existing works on generative recommendation, we found that most of them treat item tokenization and generative recommendation as two independent stages which may not be optimal for generative recommendation. In contrast, in this work, we achieved integration between item tokenization and generative recommendation via a recommendation-oriented alignment for superior recommendation performance.

\begin{figure*}
    \centering
    \includegraphics[width=0.95\linewidth]{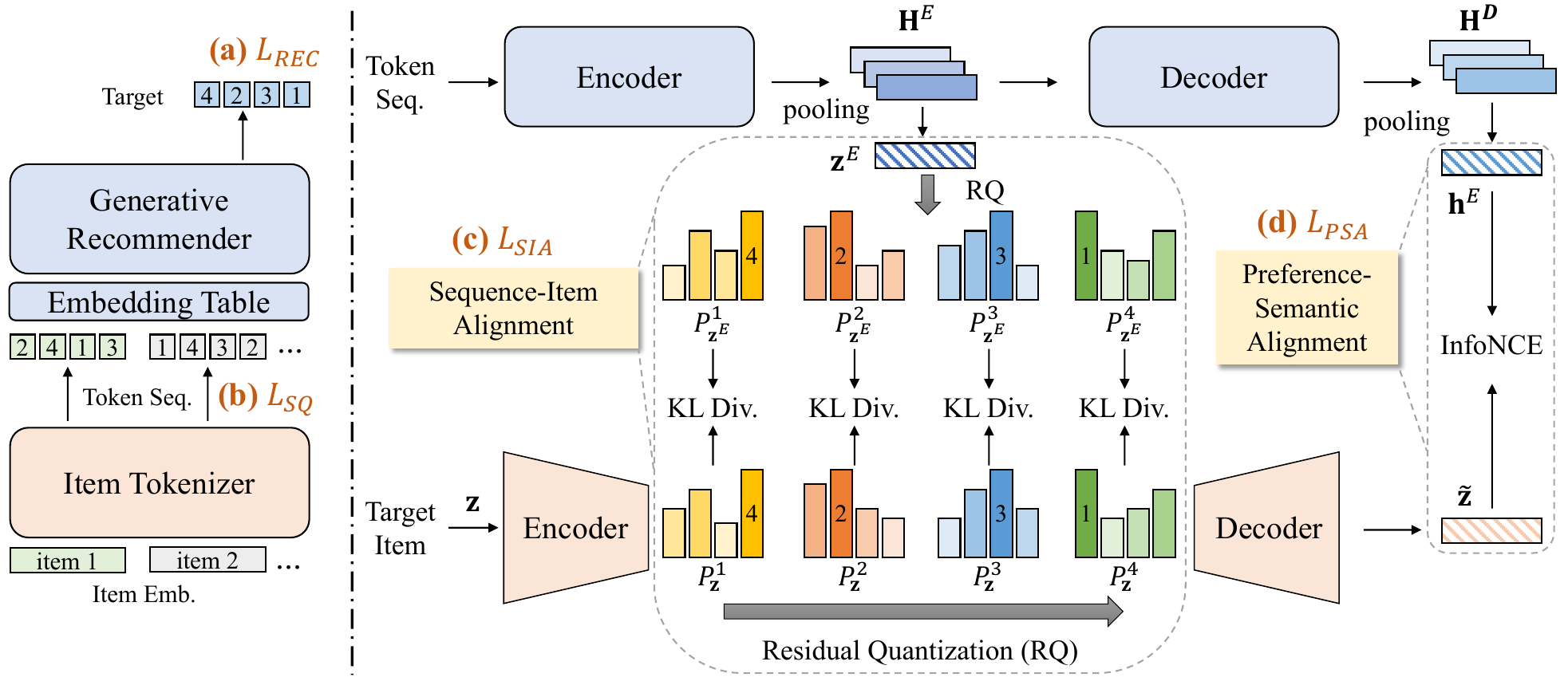}
    \caption{The overall framework of ETEGRec. ETEGRec consists of two main components, the item tokenizer and the generative recommender. Sequence-Item Alignment (SIA) and Preference-Semantic Alignment (PSA) achieve their alignment from two different perspectives for mutual enhancement.}
    \label{fig:model}
\end{figure*}

\section{Methodology}
\label{sec:methodology}
In this section, we elaborate on ETEGRec, which develops a joint optimization framework for item tokenization and generative recommendation. 
In Section~\ref{sec:prob}, we formally define the generative recommendation task. 
Then we present the dual encoder-decoder architecture of our framework in Section~\ref{sec:arch}, comprising an item tokenizer and a generative recommender. 
In Section~\ref{sec:align}, we introduce the recommendation-oriented alignment including two alignment objectives, \ie sequence-item alignment and preference-semantic alignment, to align the two components from two different perspectives. 
Finally, we describe in detail the alternating optimization approach in Section~\ref{sec:opt}. An illustration of our proposed framework is shown in Fig.~\ref{fig:model}.     

\subsection{Problem Formulation}
\label{sec:prob}
As the task setting, following prior studies~\cite{sasrec,bert4rec}, we consider a typical sequential recommendation scenario. 
Given the item set $\mathcal{I}$ and an interaction sequence $S=[i_1, i_2, \dots, i_t]$ from a target user $u$, sequential recommendation aims to predict the next item $i_{t+1} \in\mathcal{I} $ that user $u$ is likely to interact with. 
To approach this task, we take the generative paradigm by casting sequential recommendation as \emph{token sequence generation}~\cite{tiger}: each item is indexed or represented by a specific ID identifier, and the task aims to generate the ID identifier of the future interacted item.  
As a straightforward approach, we can represent each item with its original item ID~(\eg a unique product number or a randomly assigned number). However, such an approach cannot effectively share similar semantics among different items, while often leads to a large item vocabulary.

In a more generic way, each item $i$ is represented by multiple tokens $[c_1,\dots,c_L]$, where $L$ denotes the identifier length.
In practice, $L$ can vary for different items, while we follow RQ-VAE~\cite{DBLP:journals/taslp/ZeghidourLOST22} to use the same identifier length for all the items, to reduce the potential length bias in item prediction~\cite{letter,seater}.  
Following the convention in natural language processing~\cite{t5}, we refer to the process of mapping an item into multiple tokens as \emph{item tokenization}.
In this representation scheme, 
the input interaction sequence $S$ can be first tokenized into the token sequence $X=[c^1_1,c^1_2, \dots, c^t_{L-1},c^t_L]$, where each item in $S$ is represented by its  identifier (\ie $L$ tokens) and $t$ is the original item index in the interaction sequence $S$. 
To obtain the item identifiers, previous studies either {adopt the heuristic method~\cite{gpt4rec,p5,DBLP:conf/sigir-ap/HuaXGZ23} or employ a pre-learned tokenizer~\cite{tiger,letter} for item tokenization}. 
In this work, we consider devising an end-to-end approach {for learning both the item tokenizer and the recommender backbone}. 
The objective of generative recommendation is to first derive the token sequence $X$ and then generate the corresponding identifier of the target item $Y=[c^{t+1}_1,\dots,c^{t+1}_L]$ at the $(t+1)$-th step.
Formally, this task can be formulated into a typical sequence-to-sequence learning problem as follows:
\begin{equation}
P(Y|X)=\prod_{l=1}^L P(c^{t+1}_l|X,c^{t+1}_1,..,c^{t+1}_{l}).
\end{equation}

\subsection{Dual Encoder-Decoder Architecture}
\label{sec:arch}
Our proposed framework consists of two main components, namely the \emph{item tokenizer} $\mathcal{T}$ and  \emph{generative recommender}  $\mathcal{R}$,  both taking the encoder-decoder architecture. 
The input item-level interaction sequence is first mapped into the token sequence by the item tokenizer before being fed into the recommender. 
Then the generative recommender models the token sequence and autoregressively generates the item tokens for recommendation. 

\subsubsection{Item Tokenizer}
As introduced in Section~\ref{sec:prob}, we adopt a $L$-level hierarchical representation scheme for item tokenization, in which each item is indexed by $L$ token IDs. 
By taking a hierarchical scheme, we essentially organize the items in a tree-structured way, which is particularly suitable for generative tasks. 
Another merit is that the collaborative semantics among items can be shared by the same prefix tokens.    
Based on the above general idea, we next introduce the details for instanting the item tokenizer.

\paratitle{Token Generation as Residual Quantization}.
We implement the item tokenizer as a RQ-VAE, which constructs multi-level tokens via residual quantization.
For each item $i$, the item tokenizer $\mathcal{T}$ takes as input its contextual or collaborative semantic embedding $\bm{z} \in \mathbb{R}^{d_s}$\footnote{To obtain the semantic embeddings $\bm{z}$, we can run conventional recommendation algorithms (\eg SASRec) over the interaction data.}, where $d_s$ is the dimension of the semantic embedding. 
The output is its quantized tokens at each level, denoted as:
\begin{equation}
    [c_1,\dots,c_L]=\mathcal{T}(\bm{z}),
\end{equation}
where  $c_l$ denotes the corresponding token of $i$ at the $l$-th level. 
Specifically, we first encode $\bm{z}$ into a latent representation via a multilayer perceptron network~(MLP) based encoder:
\begin{equation}
    \bm{r}=\operatorname{Encoder}_T(\bm{z}).
\end{equation}
Then, the latent representation $\bm{r}$ is quantized into serialized codes (called \emph{tokens}) by looking up $L$-level codebooks, where $L$ is the token length of each item. 
At each level $l\in\{1,\dots,L\}$, we have a codebook $\mathcal{C}_l=\{\bm{e}^l_k\}_{k=1}^K,\bm{e}^l_k\in\mathbb{R}^{d_c}$, where $K$ is the codebook size.
Subsequently, the residual quantization can be conducted as:
\begin{align}
    c_l &= \arg\max_k P(k|\bm{v}_l), \\
    \bm{v}_l &= \bm{v}_{l-1} - \bm{e}^l_{c_{l-1}}.
\end{align}
where $c_l$ is the $l$-th assigned token, $\bm{v}_l$ is the residual vector at the $i$-th level, and we set $\bm{v}_1 = \bm{r}$. 
In the above equation, $P(c_l=k|\bm{v}_l)$ represents the likelihood that the residual is quantized to token $k$, which is measured by the distance between $\bm{v}_l$ and different codebook vectors.
This probability can be formulated as:
\begin{equation}
    \label{eq:dist}
    P(k|\bm{v}_{l})= \frac{\exp(-||\bm{v}_{l}-\bm{e}^l_k||^2) }{\sum^K_{j=1}\exp(-||\bm{v}_{l}-\bm{e}^l_j||^2)}.
\end{equation}

\paratitle{Reconstruction Loss}. 
Through the above process, RQ-VAE quantifies the initial semantic embedding into different levels of tokens from a coarse-to-fine granularity~\cite{DBLP:journals/taslp/ZeghidourLOST22,tiger}. 
After that, we can obtain the item tokens $[c_1,\dots,c_L]$ and the quantized representation $\tilde{\bm{r}}=\sum_{l=1}^L\bm{e}^l_{c_l}\in\mathbb{R}^{d_c}$.
Subsequently, $\tilde{\bm{r}}$ is fed into a MLP based decoder to reconstruct the item semantic embedding:
\begin{equation}
\tilde{\bm{z}}=\operatorname{Decoder}_{T}(\tilde{\bm{r}}). 
\label{eq:rq_recon}
\end{equation}
The semantic quantization loss for learning the item tokenizer is formulated as follows:
\begin{align}
\mathcal{L}_{\operatorname{SQ}}&=\mathcal{L}_{\operatorname{RECON}}+\mathcal{L}_{\operatorname{RQ}}, \label{eq:sq} \\ 
    \mathcal{L}_{\operatorname{RECON}}&=||\bm{z}-\tilde{\bm{z}}||^2, \\
    \mathcal{L}_{\operatorname{RQ}}&=\sum_{l=1}^{L}||\operatorname{sg}[\bm{v}_{l}]-\bm{e}^l_{c_l}||^2+\beta||\bm{v}_{l}-\operatorname{sg}[\bm{e}^l_{c_l}]||^2,
\end{align}   
where $\operatorname{sg}[\cdot]$ denotes the stop-gradient operation~\cite{DBLP:conf/nips/OordVK17}, and
$\beta$ is the coefficient that balances the optimization between the encoder and codebooks, typically set to 0.25.
$\mathcal{L}_{\operatorname{RECON}}$ is the reconstruction loss to guarantee that the reconstructed semantic embedding closely matches the original embedding, 
while $\mathcal{L}_{\operatorname{RQ}}$ is the RQ loss that works to minimize the distance between codebook vectors and residual vectors.

\subsubsection{Generative Recommender}
For the generative recommender, we utilize a Transformer-based encoder-decoder architecture similar to T5~\cite{t5} for sequential behavior modeling, as it has shown effectiveness in generative recommendation studies~\cite{tiger,p5,DBLP:conf/sigir-ap/HuaXGZ23}. 

\paratitle{Token-level Seq2Seq Formulation}. During training, the item-level user interaction sequence $S$ and the target item $i_{t+1}$ are first tokenized into the token-level sequence $X=[c^1_1,c^1_2, \dots, c^t_{L-1},c^t_L]$ and $Y=[c^{t+1}_{1},\dots,c^{t+1}_L]$ by $\mathcal{T}$. 
Then the corresponding token embeddings $\bm{E}^X\in\mathbb{R}^{|X|\times d_h}$ are fed into the generative recommender for user preference modeling, where $d_h$ is the hidden size of the recommender.
Formally, we begin with token sequence encoding:
\begin{align}
    \bm{H}^E=\operatorname{Encoder}_{R}(\bm{E}^X),
    \label{eq:rec_enc}
\end{align}
where $\bm{H}^E\in\mathbb{R}^{|X|\times d_h}$ is the encoded sequence representation. 
For decoding, we add a special start token ``$\operatorname{[BOS]}$'' at the beginning of $Y$ to construct the decoder input $\tilde{Y}=[\operatorname{[BOS]},c^{t+1}_1,\dots,c^{t+1}_L]$.
Then, $\bm{H}^E$ along with $\tilde{Y}$ are fed into the decoder to extract user preference representation:
\begin{align}
    \bm{H}^D=\operatorname{Decoder}_R(\bm{H}^E, \tilde{Y}),
    \label{eq:rec_dec}
\end{align}
where $\bm{H}^D\in\mathbb{R}^{(L+1)\times d_h}$ is decoder hidden states that imply user preferences over the items.

\paratitle{Recommendation Loss}. By performing inner product with the vocabulary embedding matrix $\bm{E}$, $\bm{H}^D$ is further employed to predict the target item token at each step.
Specifically, we optimize the negative log-likelihood of target tokens based on the sequence-to-sequence paradigm:
\begin{align}
    \mathcal{L}_{\operatorname{REC}}=-\sum_{j=1}^L\log P(Y_j|X,Y_{<j}),
    \label{eq:rec_loss}
\end{align}
where $Y_j$ represents the $j$-th token of target tokens, and $Y_{<j}$ denotes the tokens before $Y_j$.
In this way, the tokens of a target item will be generated autoregressively.

\subsection{Recommendation-oriented Alignment}
\label{sec:align}
In previous work~\cite{gptrec,seater,tiger,letter}, the item tokenizer and the generative recommender are treated as two separate components:  the tokenizer is often trained in the preprocessing stage to generate tokens for each item but is subsequently fixed during recommender training.
Such an approach neglects the effect of item tokenization on the generative recommendation, which cannot adaptively learn more suitable tokenizers for the corresponding recommender. 
To address this limitation, an ideal approach is to jointly learn both the item tokenizer and the generative recommender for mutual enhancement between the two components. 
For this purpose, we devise two new training strategies for aligning the two components, namely \emph{sequence-item alignment} and \emph{preference-semantic alignment}.  

\subsubsection{Sequence-Item Alignment}
 We first introduce the training strategy for sequence-item alignment. 

\paratitle{Alignment Hypothesis}. 
To align the two components, we consider the item tokenizer as an optimizable component when training the recommender. 
We employ it to obtain the corresponding token distributions for the target item based on different inputs~(See Eq.~\eqref{eq:dist}), and generate supervision signals based on the idea that two associated inputs should produce similar token distributions via the tokenizer. 
In this way, we can utilize the derived supervision signals to optimize the involved two components. 
In our approach, we adopt an encoder-decoder architecture, and assume that the hidden states $\bm{H}^E$ (Eq.~\eqref{eq:rec_enc}) from the encoder should be highly related to the collaborative embedding $\bm{z}$.
The former $\bm{H}^E$ encodes the entire information of the past interaction sequence, while the latter $\bm{z}$ captures the characteristics of the target item. 
When we feed both kinds of representations into the tokenizer, they should yield similar tokenization results. Thus, we refer to such an association relation as \emph{sequence-item alignment}.

\paratitle{Alignment Loss}. 
Based on the above alignment hypothesis, we next formulate the corresponding loss for joint optimization. Specially, we first linearize the hidden state $\bm{H}^E$ by applying a mean pooling operation: 
\begin{align}
    & \bm{z}^E = \text{MLP}(\operatorname{mean\_pool}(\bm{H}^E)),
\end{align}
where an additional MLP layer is further applied for semantic space transformation. 
Subsequently, we employ the item tokenizer to generate the token distribution for each level (Eq.~\eqref{eq:dist}), and let $P_{\bm{z}}^l$ and  $P_{\bm{z}^E}^l$ denote  
the token distributions at the $l$-th level for inputs of $\bm{z}$ (\emph{collaborative item embedding}) and $\bm{z}^E$ (\emph{encoder's sequence state}), respectively. 
Our objective is to enforce the two distributions to be similar, since the past sequence state should be highly informative for predicting the future interaction. 
Formally, we introduce the symmetric Kullback-Leibler divergence loss is as follows: 
\begin{equation}
    \label{eq:sia}
    \mathcal{L}_{\operatorname{SIA}}=-\sum_{l=1}^L\left(    D_{KL}\big(P_{\bm{z}}^l||P_{\bm{z}^E}^l)\big)+
D_{KL}\big(P_{\bm{z}^E}^l||P_{\bm{z}}^l)\big)
    \right),
\end{equation}
where $D_{KL}(\cdot)$ is the Kullback-Leibler divergence between two probablity distributions.

In addition to component fusion, another merit of this alignment loss is that it can enhance the representative capacity of the encoder. It has been found that the decoder might bypass the encoder (\ie seldom using the information from the encoder) to fulfill the generation task~\cite{DBLP:journals/corr/abs-2309-10706}, so that the encoder could not be well trained in this case. Our alignment loss can alleviate this issue and improve the overall sequence representations.

\subsubsection{Preference-Semantic Alignment}
Next, we introduce the second alignment loss. 

\paratitle{Alignment Hypothesis.} 
Specifically, we aim to leverage the connection between the decoder's first hidden state $\bm{h}^D$ (the first column in $\bm{H}^D$ from Eq.~\eqref{eq:rec_dec}) and the reconstructed semantic embedding $\tilde{\bm{z}}$  (Eq.~\eqref{eq:rq_recon}). 
The former $\bm{h}^D$ is learned by modeling the interaction sequence and reflects the sequential user preference, while the latter $\tilde{\bm{z}}$ encodes the collaborative semantics of the target item. 
Therefore, we refer to such an association relation as \emph{preference-semantic alignment}. Note that different from the recommendation loss, we use the reconstructed embedding $\tilde{\bm{z}}$, so that it naturally involves the tokenizer component in the optimization process. 

\paratitle{Alignment Loss.} Next we employ InfoNCE~\cite{DBLP:journals/jmlr/GutmannH10} with in-batch negatives to align $\bm{h}^D$ (also with MLP transformation) and $\tilde{\bm{z}}$, the  preference-semantic alignment loss is defined as follows:
\begin{align}
    \mathcal{L}_{\operatorname{PSA}}=- \left(\log\frac{\exp{(\operatorname{s}(\tilde{\bm{z}}},\bm{h}^D)/\tau)}{\sum_{\hat{\bm{h}}\in\mathcal{B}}\exp{(\operatorname{s}(\tilde{\bm{z}}},\hat{\bm{h}})/\tau)}+
    \log\frac{\exp{(\operatorname{s}(\bm{h}^D,\tilde{\bm{z}}})/\tau)}{\sum_{\hat{\bm{z}}\in\mathcal{B}}\exp{(\operatorname{s}(\bm{h}^D,\hat{\bm{z}}})/\tau)}
    \right),
    \label{eq:psa}
\end{align}
where $s(\cdot,\cdot)$ is the cosine similarity function, $\tau$ is a temperature coefficient and $\mathcal{B}$ denotes a batch of training instances.
This loss can also be considered as an additional enhancement of the recommendation loss (Eq.~\eqref{eq:rec_loss}), which uses the tokens of target items.
By incorporating the reconstructed collaborative embedding, this loss involves the tokenizer component during training, which can enhance mutual optimization. 

Through the above two alignment strategies, we can effectively enhance the association between the two components during model optimization and thus can facilitate mutual enhancement by making necessary adaptations to each other.

\subsection{Alternating Optimization}
\label{sec:opt}
Based on the dual encoder-decoder architecture and recommendation-oriented alignment, a straightforward approach is to jointly optimize the objectives of the item tokenizer and generative recommender as well as the alignment losses.
In order to improve the training stability, we propose an alternating optimization strategy to mutually train the item tokenizer and the generative recommender.

\paratitle{Item Tokenizer Optimization.}
The item tokenizer is optimized by jointly considering the semantic quantization loss $\mathcal{L}_{\operatorname{SQ}}$ (Equation~\eqref{eq:sq}), sequence-item alignment loss $\mathcal{L}_{\operatorname{SIA}}$ (Equation~\eqref{eq:sia}) and preference-semantic alignment loss $\mathcal{L}_{\operatorname{PSA}}$ (Equation~\eqref{eq:psa}),  while keeping all parameters of the generative recommender fixed. 
The overall loss can be denoted as follows:
\begin{align}
    \mathcal{L}_{\operatorname{IT}} =  \mathcal{L}_{\operatorname{SQ}}+\mu\mathcal{L}_{\operatorname{SIA}}+\lambda\mathcal{L}_{\operatorname{PSA}},
    \label{eq:token_loss}
\end{align}
where $\mu$ and $\lambda$ are hyperparameters for the trade-off of the alignment losses.

\paratitle{Generative Recommender Optimization.}
As for the generative recommender, we optimize it through the generative recommendation loss $\mathcal{L}_{\operatorname{REC}}$ (Eq.~\eqref{eq:rec_loss}), the above two alignment losses $\mathcal{L}_{\operatorname{SIA}}$ (Eq.~\eqref{eq:sia}) and $\mathcal{L}_{\operatorname{PSA}}$ (Eq.~\eqref{eq:psa}), while freezing all parameters of the item tokenizer.
The optimization objective is:
\begin{align}
    \mathcal{L}_{\operatorname{GR}} = \mathcal{L}_{\operatorname{REC}}+\mu\mathcal{L}_{\operatorname{SIA}}+\lambda\mathcal{L}_{\operatorname{PSA}}.
\end{align}

In general, we divide the training process into multiple cycles, each consisting of a fixed number of epochs. 
In the first epoch of each cycle, we optimize the item tokenizer based on Eq.~\eqref{eq:token_loss} to improve the quality of item representations by the generative recommender. 
As for the rest epochs of each cycle, the item tokenizer is frozen and item tokens remain fixed during the generative recommender training process. 
This alternation continues until the item tokenizer converges, after which we permanently freeze it and fully train the generative recommender to convergence
This approach ensures stable optimization when conducting the recommendation-oriented alignment.

\subsection{Discussion and Analysis} 
\label{sec:discussion}

\subsubsection{Comparison with Existing methods}
Recently, there have been notable advancements in generative recommendation models. 
To highlight the innovations and distinctions of our proposed approach, we conduct a comparative analysis between ETEGRec and several typical generative recommendation models from two perspectives: item tokenization and generative recommendation, as presented in Table~\ref{tab:discuss}. 

\textbf{Item tokenization} in current generative recommendation models can be broadly classified into two categories: heuristic methods and pre-learned methods~\cite{tiger,letter}. 
Heuristic methods, such as GPTRec~\cite{gptrec} and CID~\cite{DBLP:conf/sigir-ap/HuaXGZ23}, employ manually constructed user-item interaction matrix or item co-occurrence matrix to estimate similarity between items.
Although these methods are straightforward and efficient, they often fail to capture the profound semantic relevance between items.
As for pre-learned methods like TIGER and LETTER, they pre-learn a deep neural network as the item tokenizer (\eg autoencoder) to derive identifiers with implicit semantics.
However, these methods treat item tokenization as a preprocessing step, resulting in a \emph{complete decoupling} of item tokenizer and generative recommender during model optimization.
In contrast, ETEGRec \emph{integrates the tokenizer and recommender into an end-to-end framework} to address this decoupling problem and achieves mutual enhancement between the two components by proposing a recommendation-oriented alignment approach.
Furthermore, from the interaction-aware perspective, only GPTRec introduces interaction awareness through the user-item interaction matrix.
Different from them, ETEGRec aligns the past user interaction sequence and the target item from two different perspectives, thereby incorporating the preference information within user behaviors into the item tokenizer.

\textbf{Generative recommendation} in existing methods typically processes user interaction sequences into corresponding token sequences in advance. 
Such constant data suffer from \emph{monotonous} sequence patterns, which brings the risk of overfitting.
In contrast, ETEGRec jointly optimizes the item tokenizer during model learning, resulting in \emph{diverse} token sequences and \emph{gradually refined} semantics.
The ablation experiment in Section~\ref{sec:ablation} confirmed that the continuous enhancement of token sequences significantly contributes to the performance.
Moreover, unlike existing methods that isolate the item tokenizer in generative recommendation, our approach further integrates and refines the prior knowledge implicit in item semantic embeddings from the item tokenizer.

\begin{table}
\centering
\caption{
Comparison of ETEGRec with several related studies on item Tokenization and generative recommendation.
``EL'' means the length of item identifiers are equal. ``IA'' denotes interaction-aware. ``TI'' denotes tokenization integration. 
}
\label{tab:discuss}
\resizebox{\linewidth}{!}{
\begin{tabular}{lcccccc}
\toprule
\multirow{2}{*}{Methods}  & \multicolumn{3}{c}{Item Tokenization} & \multicolumn{2}{c}{Generative Recommendation} \\
\cmidrule(l){2-4} \cmidrule(l){5-6}
 & Learning & EL &IA &  Token Sequence & TI  \\
\midrule
GPTRec~\cite{gptrec} &Heuristic & \CheckmarkBold & \CheckmarkBold & Pre-processed & \XSolidBrush  \\
CID~\cite{DBLP:conf/sigir-ap/HuaXGZ23} &Heuristic & \XSolidBrush & \XSolidBrush & Pre-processed  & \XSolidBrush \\
TIGER~\cite{tiger} & Pre-learned & \CheckmarkBold & \XSolidBrush & Pre-processed & \XSolidBrush   \\
LETTER~\cite{letter} & Pre-learned & \CheckmarkBold & \XSolidBrush & Pre-processed & \XSolidBrush  \\
ETEGRec  & End-to-end & \CheckmarkBold  & \CheckmarkBold & Gradually Refined & \CheckmarkBold   \\
\bottomrule
\end{tabular}}
\end{table}

\subsubsection{Complexity Analysis}
For item tokenization, considering a single item as an example, the time complexity of the encoder and decoder layers is $\mathcal{O}(d^2)$, where $d$ is the model dimension. The time complexity for codebook lookup operations is $\mathcal{O}(LKd)$, where $K$ is the size per codebook and $L$ is the codebook number. The computation complexity of the semantic quantization loss is $\mathcal{O}(d+Ld)$. Thus, the total time complexity for tokenizing one item is $\mathcal{O}(d^2+LKd)$. For generative recommendation, the time complexity of sequential preference modeling primarily stems from the self-attention and feed-forward layers, which is $\mathcal{O}(N^2d+Nd^2)$ where $N$ represents the sequence length. The time used by calculating losses of REC, SIA and PSA is $\mathcal{O}(LKd)$, $\mathcal{O}({LKd})$ and $\mathcal{O}(Md)$, respectively, where $M$ is the number of negative samples. Overall, the training cost is $\mathcal{O}(NLKd+N^2d+Nd^2+Md)$, which is on the same order of magnitude as mainstream models (\eg TIGER and LETTER).
The inference complexity of our proposed method is completely consistent with TIGER, as the item tokenization results can be cached in advance.
\section{Experiments}
\label{sec:experiments}

\begin{table}[]
    \centering
    \caption{Statistics of the Datasets.}
    \begin{tabular}{ccccc}
    \toprule
     Dataset    &\#Users   &\#Items   &\#Interactions &Sparsity  \\
     \midrule
     Instrument &57,439  &24,587  &511,836  &99.964\% \\
     Scientific &50,985  &25,848  &412,947  &99.969\% \\
     Game &94,762  &25,612  &814,586  &99.966\% \\
     
     \bottomrule
    \end{tabular}
    \label{tab:data_statistics}
\end{table}

In this section, we begin with the detailed experiment setup and then present overall performance and in-depth analysis of our proposed approach.

\begin{table*}[]
\centering
\caption{The overall performance comparisons between the baselines and ETEGRec. The best and second-best results are highlighted in bold and underlined font, respectively. * denotes that the improvements are statistically significant with $p<0.01$ in a paired t-test setting.}
\label{tab:main_result}
\huge
\resizebox{\textwidth}{!}{
\renewcommand\arraystretch{1.15}
\setlength{\tabcolsep}{0.65mm}{
\begin{tabular}{l|cccc|cccc|cccc}
\hline
\multirow{2}{*}{Model} & \multicolumn{4}{c|}{Instrument}                                    & \multicolumn{4}{c|}{Scientific}                                          & \multicolumn{4}{c}{Game}                                           \\
\cline{2-13}                      & Recall@5 & Recall@10 & NDCG@5 & NDCG@10 & Recall@5 & Recall@10 & NDCG@5 & NDCG@10 & Recall@5 & Recall@10 & NDCG@5 & NDCG@10 \\ \hline
Caser      & 0.0242   & 0.0392    & 0.0154 & 0.0202  & 0.0172   & 0.0281    & 0.0107 & 0.0142  & 0.0346   & 0.0567    & 0.0221 & 0.0291  \\
GRU4Rec                                    & 0.0345   & 0.0537    & 0.0220  & 0.0281  & 0.0221   & 0.0353    & 0.0144 & 0.0186  & 0.0522   & 0.0831    & 0.0337 & 0.0436  \\
HGN                                        & 0.0319   & 0.0515    & 0.0202 & 0.0265  & 0.0220    & 0.0356    & 0.0138 & 0.0182  & 0.0423   & 0.0694    & 0.0266 & 0.0353  \\
SASRec                                     & 0.0341   & 0.0530     & 0.0217 & 0.0277  & 0.0256   & 0.0406    & 0.0147 & 0.0195  & 0.0517   & 0.0821    & 0.0329 & 0.0426  \\
BERT4Rec                                   & 0.0305   & 0.0483    & 0.0196 & 0.0253  & 0.0180    & 0.0300      & 0.0113 & 0.0151  & 0.0453   & 0.0716    & 0.0294 & 0.0378  \\
FMLP-Rec                                   & 0.0328   & 0.0529    & 0.0206 & 0.0271  & 0.0248   & 0.0388    & 0.0158 & 0.0203  & 0.0535   & 0.0860     & 0.0331 & 0.0435  \\
FDSA                                       & 0.0364   & 0.0557    & 0.0233 & 0.0295  & 0.0261   & 0.0391    & 0.0174 & 0.0216  & 0.0548   & 0.0857    & 0.0353 & 0.0453  \\
S$^3$Rec                                      & 0.0340    & 0.0538    & 0.0218 & 0.0282  & 0.0253   & 0.0410    & 0.0172 & 0.0218  & 0.0533   & 0.0823    & 0.0351 & 0.0444  \\ \hline
SID                                     & 0.0319   & 0.0438    & 0.0237 & 0.0275  & 0.0155   & 0.0234    & 0.0103 & 0.0129  & 0.0480    & 0.0693    & 0.0333 & 0.0401  \\
CID                                     & 0.0352   & 0.0507    & 0.0234 & 0.0285  & 0.0192   & 0.0300      & 0.0123 & 0.0158  & 0.0497   & 0.0748    & 0.0343 & 0.0424  \\
TIGER                                      & 0.0368          & 0.0574          & {0.0242}   & {0.0308}    & {0.0275}    & 0.0431          & \ul{0.0181}   & \ul{0.0231}    & {0.0570}    & {0.0895}    & {0.0370}  & {0.0471} \\
TIGER-SAS                                  & \ul{0.0375}    & 0.0576   & 0.0242   & 0.0306          & 0.0272          & \ul{0.0435}    & 0.0174         & 0.0227          & 0.0561          & 0.0891          & 0.0363        & 0.0469  \\
LETTER                                  & 0.0372    & \ul{0.0581}    & \ul{0.0243}   & \ul{0.0310}          & \ul{0.0276}          & 0.0433    & 0.0179         & 0.0230          & \ul{0.0576}          & \ul{0.0901}          & \ul{0.0373}        & \ul{0.0475}  \\ \hline
ETEGRec                                    & \textbf{0.0402}$^*$ & \textbf{0.0624}$^*$ & \textbf{0.0260}$^*$ & \textbf{0.0331}$^*$ & \textbf{0.0294}$^*$ & \textbf{0.0455}$^*$ & \textbf{0.0190}$^*$ & \textbf{0.0241}$^*$ & \textbf{0.0616}$^*$ & \textbf{0.0947}$^*$ & \textbf{0.0400}$^*$ & \textbf{0.0507}$^*$ \\ \hline

\end{tabular}}}
\end{table*}

\subsection{Experiment Setup}

\subsubsection{Dataset}

We conduct experiments on three subsets of the most recent Amazon 2023 review data~\cite{DBLP:journals/corr/abs-2403-03952} to evaluate our approach, including ``\emph{Musical Instruments}'', ``\emph{Video Games}'', and ``\emph{Industrial Scientific}''.
All these datasets comprise user review data from May 1996 to September 2023. 
Following previous works~\cite{s3rec,lcrec}, we apply the 5-core filter to exclude unpopular users and items with less than five interaction records. 
Then, we construct user behavior sequences according to the chronological order and uniformly set the maximum item sequence length to 50.
The statistics of preprocessed datasets are shown in Table~\ref{tab:data_statistics}.

\subsubsection{Baseline Models}
The baseline models we adopt for comparison include the following two categories:
\noindent (1) \emph{Traditional sequential recommendation models}:
{\textbf{Caser}}~\cite{DBLP:conf/wsdm/TangW18} utilizes horizontal and vertical convolutional filters to model user behavior sequences.
{\textbf{HGN}}~\cite{DBLP:conf/kdd/MaKL19} employs hierarchical gating networks to capture both long-term and short-term user interests from item sequences.
{\textbf{GRU4Rec}}~\cite{DBLP:journals/corr/HidasiKBT15} is an RNN-based sequential recommender that uses GRU for user behavior modeling.
{\textbf{BERT4Rec}}~\cite{bert4rec} introduces bidirectional Transformer and mask prediction tasks into sequential recommendation for user preference modeling.
{\textbf{SASRec}}~\cite{sasrec} adopts the unidirectional Transformer to model user behaviors and predict the next item.
{\textbf{FMLP-Rec}}~\cite{DBLP:conf/www/ZhouYZW22} proposes an all-MLP sequential recommender with learnable filters, which can effectively reduce user behavior noise.
{\textbf{FDSA}}~\cite{fdsa} emphasizes the transformation patterns between item features by separately modeling both item-level and feature-level sequences using self-attention networks.
{\textbf{S$^3$-Rec}}~\cite{s3rec} incorporates mutual information maximization into sequential recommendation for model pre-training, learning the correlation between items and attributes to improve recommendation performance.
\noindent (2) \emph{Generative recommendation models}:
{\textbf{SID}}~\cite{DBLP:conf/sigir-ap/HuaXGZ23} sequentially encodes item IDs according to user interaction order and utilizes them as item identifiers for generative recommendation.
{\textbf{CID}}~\cite{DBLP:conf/sigir-ap/HuaXGZ23} integrates collaborative knowledge into LLM-based generative recommender by generating item identifiers through spectral clustering on item co-occurrence graphs.
{\textbf{TIGER}}~\cite{tiger} leverages text embedding to construct semantic IDs for items and adopts the generative retrieval paradigm for sequential recommendation.
{\textbf{TIGER-SAS}}~\cite{tiger} uses the item embeddings from trained SASRec instead of text embeddings to construct semantic IDs, which enables item identifiers to imply collaborative prior knowledge.
{\textbf{LETTER}}~\cite{letter} designs a learnable tokenizer by integrating hierarchical semantics, collaborative signals, and code assignment diversity.

\begin{table*}[]
\centering
\caption{Ablation study of ETEGRec. We assess the proposed two alignment objectives and the alternating training strategy.}
\label{tab:ablation}
\resizebox{\textwidth}{!}{
\renewcommand\arraystretch{1.15}
\setlength{\tabcolsep}{0.6mm}{
\begin{tabular}{l|cccc|cccc|cccc}
\hline
\multirow{2}{*}{Variants}                & \multicolumn{4}{c|}{{Instrument}} & \multicolumn{4}{c|}{{Scientific}} & \multicolumn{4}{c}{{Game}}     \\ \cline{2-13}
 & {Recall@5} & {Recall@10}  & {NDCG@5} & {NDCG@10}  & {Recall@5} &{Recall@10} & {NDCG@5} &{NDCG@10} & {Recall@5} &{Recall@10} & {NDCG@5} &{NDCG@10} \\ \hline
ETEGRec  & \textbf{0.0402} & \textbf{0.0624} & \textbf{0.0260} & \textbf{0.0331} & \textbf{0.0294} & \textbf{0.0455} & \textbf{0.0190} & \textbf{0.0241} & \textbf{0.0616} & \textbf{0.0947} & \textbf{0.0400} & \textbf{0.0507}           \\
 \ \ \emph{w/o} $\mathcal{L}_{\operatorname{SIA}}$     & 0.0396          & 0.0614          & 0.0255          & 0.0325          & 0.0285          & 0.0446          & 0.0186          & 0.0238          & 0.0590           & 0.0917          & 0.0386          & 0.0491                \\
 \ \ \emph{w/o} $\mathcal{L}_{\operatorname{PSA}}$     & 0.0389          & 0.0609          & 0.0250          & 0.0321          & 0.0270          & 0.0422          & 0.0174          & 0.0223          & 0.0602          & 0.0933          & 0.0392          & 0.0499         \\
 \ \ \emph{w/o} $\mathcal{L}_{\operatorname{SIA}}$ \& $\mathcal{L}_{\operatorname{PSA}}$    & 0.0379          & 0.0601          & 0.0245          & 0.0317          & 0.0269          & 0.0422          & 0.0175          & 0.0224          & 0.0576          & 0.0894          & 0.0375          & 0.0478                 \\
 \ \ \emph{w/o} AT       & 0.0337          & 0.0529          & 0.0215          & 0.0277          & 0.0234          & 0.0375          & 0.0153          & 0.0198          & 0.0514          & 0.0810          & 0.0333          & 0.0428       \\
 \ \ \emph{w/o} ETE       & 0.0388          & 0.0600          & 0.0252          & 0.0320          & 0.0277          & 0.0431          & 0.0181          & 0.0230          & 0.0569          & 0.0899          & 0.0369          & 0.0475      \\
\hline
\end{tabular}}}
\end{table*}

\subsubsection{Evaluation Settings}
To evaluate the performance of various methods in sequential recommendation, we employ two widely used metrics: top-$K$ Recall and top-$K$ Normalized Discounted Cumulative Gain (NDCG), where $K$ is set to 5, and 10. 
Following prior studies~\cite{s3rec,tiger}, we employ the \emph{leave-one-out} strategy to split training, validation, and test sets. 
Specifically, for each user, the latest interaction is used as testing data, the second most recent interaction is validation data, and all other interaction records are used for training.
We conduct the full ranking evaluation over the entire item set to avoid bias introduced by sampling. 
The beam size is uniformly set to 20 for all generative recommendation models.

\subsubsection{Semantic ID Generation}
We obtain 256-dimensional item collaborative semantic embeddings from a trained SASRec~\cite{sasrec}. For the item tokenizer, we utilize a 3-layer MLP for both the encoder and decoder in the RQVAE. The number of codebooks $L$ is set to 3, with each codebook containing $K=256$ code embeddings of dimension 128. To ensure the uniqueness of semantic item IDs, following the approach in TIGER~\cite{tiger} we append an additional token at the end of the semantic tokens.

\subsubsection{Implementation Details}
For our generative recommender, we employ T5 with 6 encoder and decoder layers as the backbone. The hidden size and the dimension of the feed-forward network (FFN) are set to 128 and 512, respectively. Each layer consists of 4 self-attention heads with a dimension of 64. We utilize a pre-trained RQ-VAE to initialize our item tokenizer and employ the AdamW optimizer with a weight decay of 0.05 to train the entire framework. The number of epochs per cycle $C$ is tuned in \{2, 4\}. The training process begins with training the item tokenizer for 1 epoch, followed by training the generative recommender for $C$-1 epochs. This process is repeated until convergence based on validation performance. The learning rates for the generative recommender and item tokenizer are tuned within the ranges of \{5e-3, 3e-3, 1e-3\} and \{5e-4, 1e-4, 5e-5\}, respectively. The hyper-parameters $\mu$ and $\lambda$ are tuned within the range \{5e-3, 1e-3, 5e-4, 3e-4, 1e-4\}.

Traditional recommendation baselines are implemented by an open-source recommendation framework RecBole~\cite{recbole1.0,recbole2.0}. For CID, SID~\cite{DBLP:conf/sigir-ap/HuaXGZ23} and LETTER~\cite{letter}, we utilize their official implementations. For TIGER and TIGER-SAS, we follow the implementation details provided in the original paper~\cite{tiger}. To ensure fair comparisons, the item embedding dimension is set to 128 for all models, except for SID and CID, which retain their default dimension of 768.

\subsection{Overall Performance}

We evaluate ETEGRec on three public recommendation benchmarks. The overall results are presented in Table~\ref{tab:main_result}, from which we have the following observations:

$\bullet$ Among traditional sequential recommendation models, FDSA exhibits superior performance compared with others across three datasets, which is attributed to the utilization of additional textual feature embeddings. FMLP-Rec achieves comparable performance as SASRec and BERT4Rec which suggests that all-MLP architectures can also model the behavior sequence effectively.

$\bullet$ For generative recommendation models, TIGER and TIGER-SAS consistently outperform CID and SID across all three datasets, even though CID and SID adopt the pretrained T5 model with more parameters. This performance disparity can be attributed to their distinct item tokenization methods. 
Specifically, SID leverages numerical tokens to index items, which lack semantic information. CID employs a heuristic tokenizer based on the item co-occurrence graph to construct item identifiers; however, this approach fails to effectively capture the similarity between items. In contrast, TIGER and TIGER-SAS learn hierarchical textual or collaborative semantics from coarse to fine granularity through RQ-VAE, which is more beneficial for recommendation tasks. Notably, TIGER-SAS demonstrates comparable performance to TIGER on all datasets, indicating that both collaborative and textual semantics contribute significantly to recommendation performance.
LETTER performs best across most cases, as it effectively incorporates collaborative and textual semantic information.

$\bullet$ ETEGRec consistently achieves the best results on all datasets compared to the baseline methods, which demonstrates its effectiveness. We attribute the improvements to the mutual enhancement between the item tokenizer and the generative recommender through the recommendation-oriented alignment.

\subsection{Ablation Study}
\label{sec:ablation}

In order to evaluate the impact of the proposed techniques in ETEGRec, we conduct an ablation study on all three datasets. The performance of the four variants is depicted in Table~\ref{tab:ablation}.

$\bullet$ \underline{\emph{w/o} $\mathcal{L}_{\operatorname{SIA}}$} without the sequence-item alignment (SIA) (Eq.~\eqref{eq:sia}). We can see that this variant performs worse than ETEGRec across all datasets, which indicates that alignment between sequence representation and item representation in the codebook space is beneficial for generative recommendation.

$\bullet$ \underline{\emph{w/o} $\mathcal{L}_{\operatorname{PSA}}$} removes the preference-semantic alignment (PSA) (Eq.~\eqref{eq:psa}), which also brings a performance degradation. 
The phenomenon demonstrates the effectiveness of the proposed PSA loss, which can enhance user preference modeling.

$\bullet$ \underline{\emph{w/o} $\mathcal{L}_{\operatorname{SIA}}$ \& $\mathcal{L}_{\operatorname{PSA}}$} without both $\mathcal{L}_{\operatorname{SIA}}$ and $\mathcal{L}_{\operatorname{PSA}}$.
The variant lacking both alignments performs worse than removing just one. These results show that both sequence-item alignment and preference-semantic alignment positively contribute to generative recommendation, with their combination leading to improved performance.

$\bullet$ \underline{\emph{w/o} AT} directly jointly learns all involved optimization objectives in our framework.
We can find that omitting the alternating training strategy from ETEGRec leads to a significant performance decline. 
This result suggests that frequent updates to the item tokenizer during training adversely affect the recommender’s training. 
By employing alternating training, we achieve stable and effective training for both components while maintaining collaborative alignment within them.

$\bullet$ \underline{\emph{w/o} ETE} bypasses the end-to-end optimization process and instead leverages the final item tokens obtained by our ETEGRec to retrain a generative recommender.
From the results, it can be seen that the improvement of ETEGRec is not only due to superior item identifiers but also attributed to the integration of prior knowledge encoded in the item tokenizer with the generative recommender.

\subsection{Further Analysis}

\begin{figure}[]
    \centering
    \includegraphics[width=0.97\linewidth]{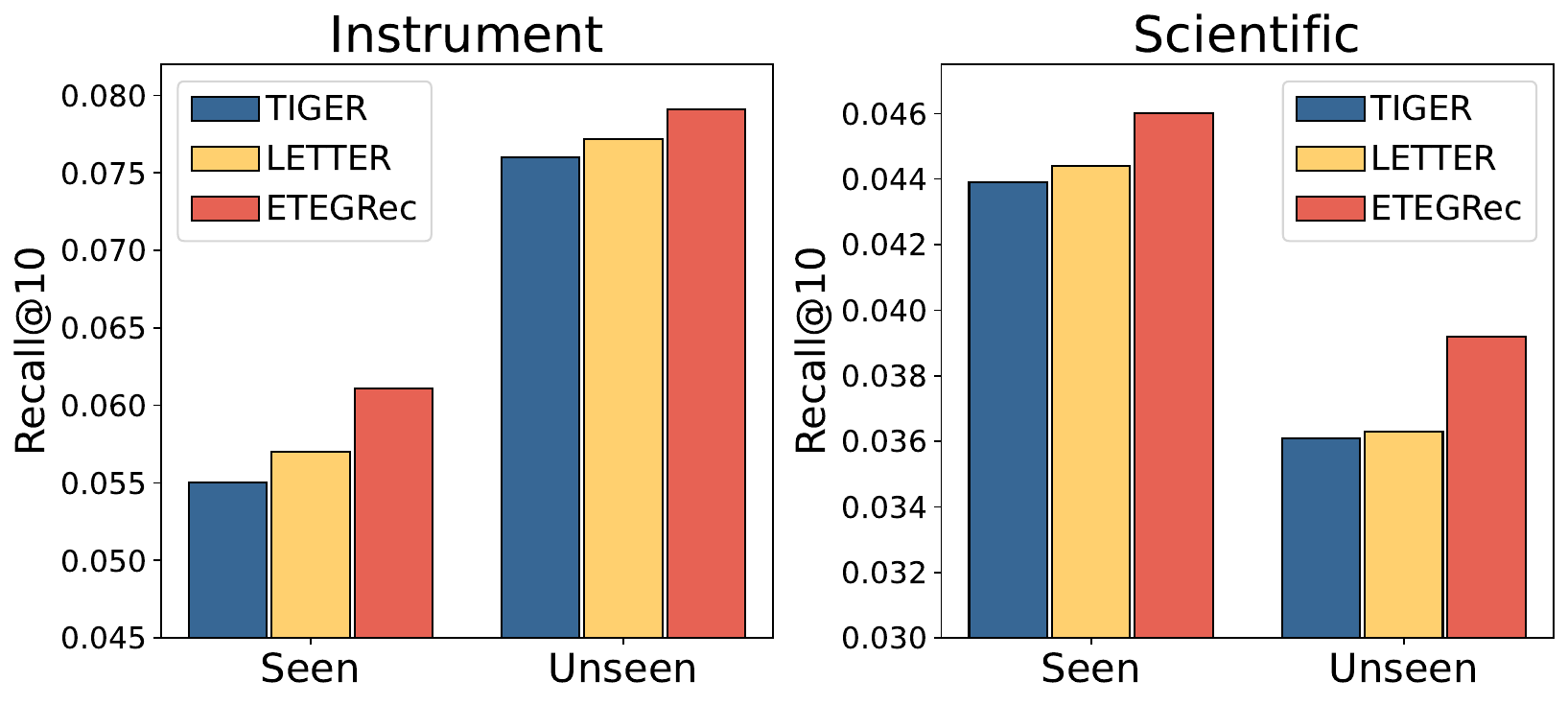}
    \caption{Performance comparison on seen and unseen users.}
    \label{fig:unseen}
\end{figure}

\subsubsection{Generalizability Evaluation}
To assess the generalizability of ETEGRec, we evaluate its recommendation performance on new users who are unseen during training.
We construct a new training set by removing the interactions of several users from the training set, and obtain a test set containing both seen and unseen users.
Specifically, we select 5\% of users with the least interaction history as new users on Instrument and Scientific datasets, and then evaluate the recommendation performance for both seen and unseen users.
From Fig.~\ref{fig:unseen}, it is evident that ETEGRec outperforms LETTER and TIGER on both seen and unseen users.
This indicates that ETEGRec processes a more robust ability to model users' preferences through the alignment between item tokenizer and generative recommender.

\subsubsection{Preference-Semantic Representation Visualization}

\begin{figure}[]
    \centering
    \subfloat[Instrument]{\includegraphics[width=0.5\linewidth]{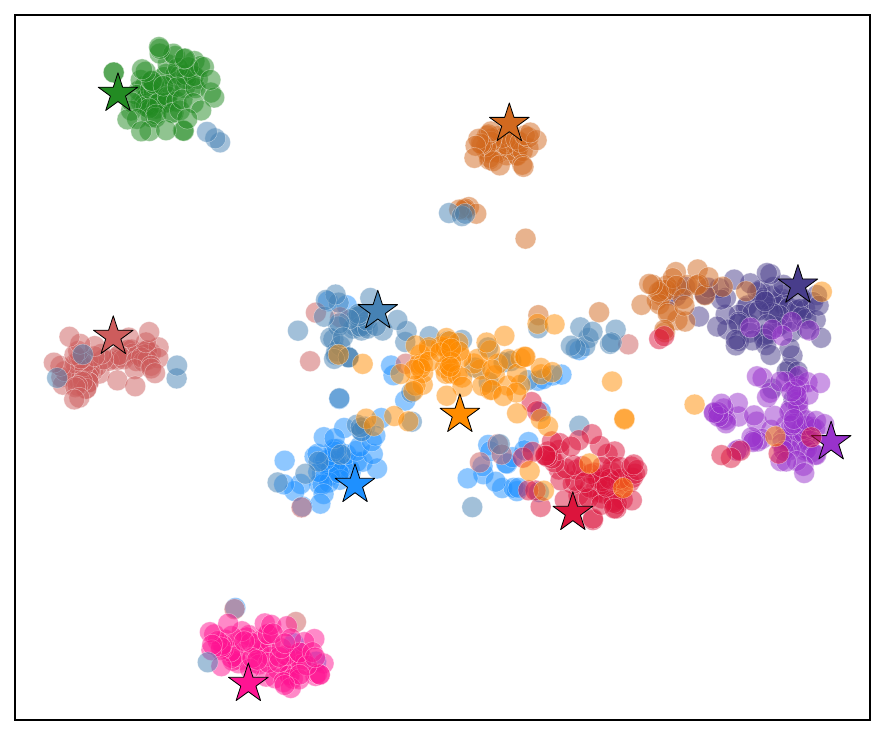}}
    \subfloat[Scientific]{\includegraphics[width=0.5\linewidth]{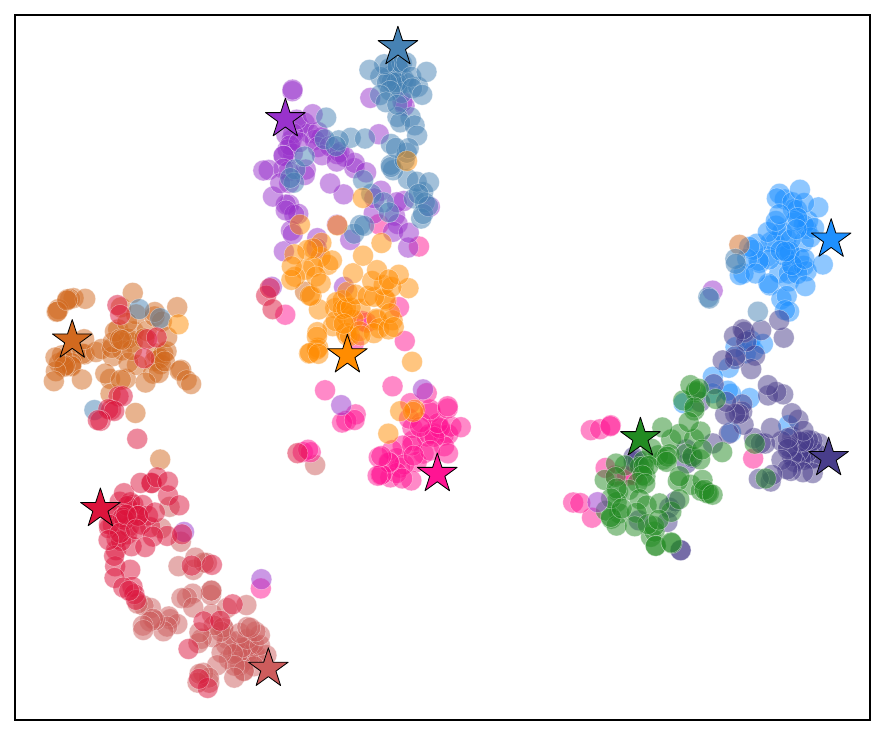}}
    \caption{Visualization of preference and semantic representations, where \textit{circles} denote preference points, \textit{stars} represent semantic points, and different colors indicate distinct groups}
    \label{fig:cluster}
\end{figure}

To further validate the effectiveness of our proposed \textit{preference-semantic alignment} (\ie Equation \eqref{eq:psa}), we employ t-SNE~\cite{tsne} to visualize the preference representation $\mathbf{h}^D$ and semantic representations of the corresponding target items $\tilde{z}$, as shown in Figure \ref{fig:cluster}. Specifically, we select 10 items and 80 corresponding interaction histories from Instrument and Scientific datasets and extract the preference and reconstructed semantic representations. From Figure \ref{fig:cluster}, we observe that the preference points are closely clustered around their corresponding target semantic points while being separated from other semantic points, demonstrating the effectiveness of our proposed PSA in aligning sequential user preference and the target item semantics.

\subsubsection{Hyper-Parameter Analysis}
 
\begin{figure}[htbp]
    \centering
    \subfloat[Instrument]{\includegraphics[width=\linewidth]{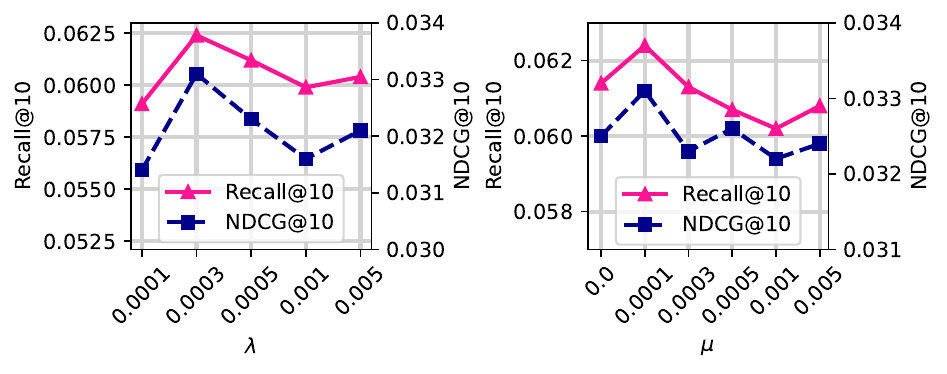}}
    \vspace{5px}
    \subfloat[Scientific]{\includegraphics[width=\linewidth]{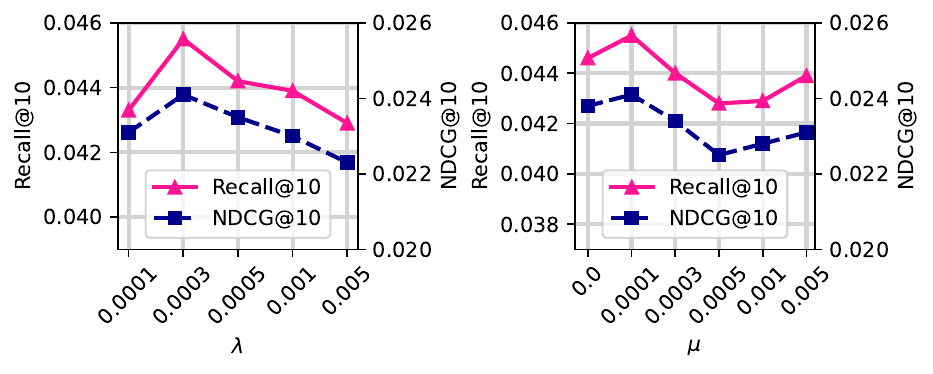}}
    \vspace{5px}
    \subfloat[Game]{\includegraphics[width=\linewidth]{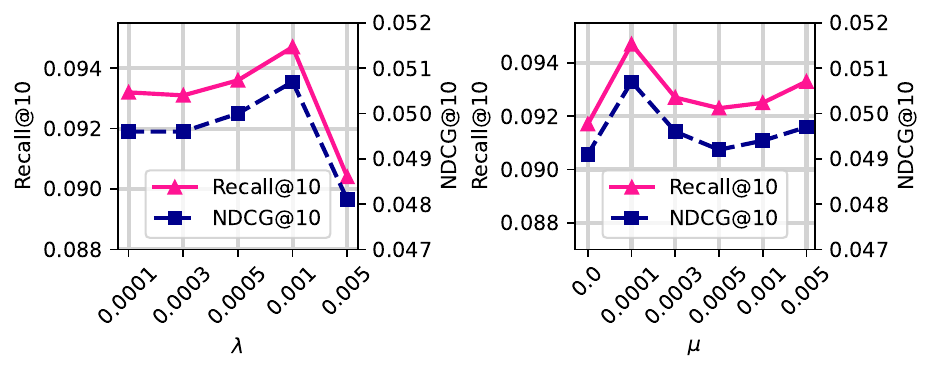}}
    \caption{Performance comparison of different alignment loss coefficients.}
    \label{fig:hyper}
\end{figure}

For sequence-item alignment, we investigate it by varying the coefficient $\mu$ from 1e-4 to 5e-3.
As illustrated in Fig.~\ref{fig:hyper}, increasing $\mu$ beyond this optimal range could interfere with model learning and adversely affect performance. The optimal results are achieved with $\mu=\text{3e-4}$ for the Instrument and Scientific dataset, and $\mu=\text{1e-3}$ for the Game dataset.
To explore the influence of preference-semantic alignment, we tune $\lambda$ within the range from 0 to 5e-3 and observe similar trends to those seen with $\mu$, as shown in Fig.~\ref{fig:hyper}.
ETEGRec yields suboptimal performance at too large $\lambda$ and performs best on all three datasets when $\lambda=\text{1e-4}$.

\section{Conclusion}
\label{sec:conclusion}

In this paper, we proposed ETEGRec, a novel end-to-end generative recommender with recommendation-oriented alignment. Different from previous methods decoupling item tokenization and generative recommendation, ETEGRec seamlessly integrated the item tokenizer and the generative recommender to build a fully end-to-end generative recommendation framework. We further designed a recommendation-oriented alignment approach, comprising \emph{sequence-item alignment} and \emph{preference-semantic alignment}, to achieve mutual enhancement of the two components from two different perspectives. To enable effective end-to-end learning, we further proposed an alternating optimization strategy for joint component learning. Extensive experiments and in-depth analysis on three benchmarks have demonstrated the superiority of our proposed framework, ETEGRec, compared to both traditional sequential recommendation models and generative recommendation baselines. In future work, we will transfer the joint tokenization method to other generative recommendation architectures, and also explore the scaling effect when increasing the model parameters. 


\begin{acks}
This work was partially supported by National Natural Science Foundation of China under Grant No. 92470205 and 62222215, Beijing Municipal Science and Technology Project under Grant No. Z231100010323009, and Beijing Natural Science Foundation under Grant No. L233008. Xin Zhao is the corresponding author.
\end{acks}

\bibliographystyle{_acm/ACM-Reference-Format}
\balance
\bibliography{bibliography}


\begin{thebibliography}{43}


\ifx \showCODEN    \undefined \def \showCODEN     #1{\unskip}     \fi
\ifx \showISBNx    \undefined \def \showISBNx     #1{\unskip}     \fi
\ifx \showISBNxiii \undefined \def \showISBNxiii  #1{\unskip}     \fi
\ifx \showISSN     \undefined \def \showISSN      #1{\unskip}     \fi
\ifx \showLCCN     \undefined \def \showLCCN      #1{\unskip}     \fi
\ifx \shownote     \undefined \def \shownote      #1{#1}          \fi
\ifx \showarticletitle \undefined \def \showarticletitle #1{#1}   \fi
\ifx \showURL      \undefined \def \showURL       {\relax}        \fi
\providecommand\bibfield[2]{#2}
\providecommand\bibinfo[2]{#2}
\providecommand\natexlab[1]{#1}
\providecommand\showeprint[2][]{arXiv:#2}

\bibitem[Chang et~al\mbox{.}(2021)]%
        {DBLP:conf/sigir/ChangGZHNSJ021}
\bibfield{author}{\bibinfo{person}{Jianxin Chang}, \bibinfo{person}{Chen Gao}, \bibinfo{person}{Yu Zheng}, \bibinfo{person}{Yiqun Hui}, \bibinfo{person}{Yanan Niu}, \bibinfo{person}{Yang Song}, \bibinfo{person}{Depeng Jin}, {and} \bibinfo{person}{Yong Li}.} \bibinfo{year}{2021}\natexlab{}.
\newblock \showarticletitle{Sequential Recommendation with Graph Neural Networks}. In \bibinfo{booktitle}{\emph{{SIGIR} '21: The 44th International {ACM} {SIGIR} Conference on Research and Development in Information Retrieval, Virtual Event, Canada, July 11-15, 2021}}, \bibfield{editor}{\bibinfo{person}{Fernando Diaz}, \bibinfo{person}{Chirag Shah}, \bibinfo{person}{Torsten Suel}, \bibinfo{person}{Pablo Castells}, \bibinfo{person}{Rosie Jones}, {and} \bibinfo{person}{Tetsuya Sakai}} (Eds.). \bibinfo{publisher}{{ACM}}, \bibinfo{pages}{378--387}.
\newblock
\href{https://doi.org/10.1145/3404835.3462968}{doi:\nolinkurl{10.1145/3404835.3462968}}


\bibitem[{Di Palma}(2023)]%
        {DBLP:conf/recsys/Palma23}
\bibfield{author}{\bibinfo{person}{Dario {Di Palma}}.} \bibinfo{year}{2023}\natexlab{}.
\newblock \showarticletitle{Retrieval-augmented Recommender System: Enhancing Recommender Systems with Large Language Models}. In \bibinfo{booktitle}{\emph{Proceedings of the 17th {ACM} Conference on Recommender Systems, RecSys 2023, Singapore, Singapore, September 18-22, 2023}}, \bibfield{editor}{\bibinfo{person}{Jie Zhang}, \bibinfo{person}{Li~Chen}, \bibinfo{person}{Shlomo Berkovsky}, \bibinfo{person}{Min Zhang}, \bibinfo{person}{Tommaso~Di Noia}, \bibinfo{person}{Justin Basilico}, \bibinfo{person}{Luiz Pizzato}, {and} \bibinfo{person}{Yang Song}} (Eds.). \bibinfo{publisher}{{ACM}}, \bibinfo{pages}{1369--1373}.
\newblock
\href{https://doi.org/10.1145/3604915.3608889}{doi:\nolinkurl{10.1145/3604915.3608889}}


\bibitem[Geng et~al\mbox{.}(2022)]%
        {p5}
\bibfield{author}{\bibinfo{person}{Shijie Geng}, \bibinfo{person}{Shuchang Liu}, \bibinfo{person}{Zuohui Fu}, \bibinfo{person}{Yingqiang Ge}, {and} \bibinfo{person}{Yongfeng Zhang}.} \bibinfo{year}{2022}\natexlab{}.
\newblock \showarticletitle{Recommendation as Language Processing {(RLP):} {A} Unified Pretrain, Personalized Prompt {\&} Predict Paradigm {(P5)}}. In \bibinfo{booktitle}{\emph{RecSys '22: Sixteenth {ACM} Conference on Recommender Systems, Seattle, WA, USA, September 18 - 23, 2022}}, \bibfield{editor}{\bibinfo{person}{Jennifer Golbeck}, \bibinfo{person}{F.~Maxwell Harper}, \bibinfo{person}{Vanessa Murdock}, \bibinfo{person}{Michael~D. Ekstrand}, \bibinfo{person}{Bracha Shapira}, \bibinfo{person}{Justin Basilico}, \bibinfo{person}{Keld~T. Lundgaard}, {and} \bibinfo{person}{Even Oldridge}} (Eds.). \bibinfo{publisher}{{ACM}}, \bibinfo{pages}{299--315}.
\newblock
\href{https://doi.org/10.1145/3523227.3546767}{doi:\nolinkurl{10.1145/3523227.3546767}}


\bibitem[Gutmann and Hyv{\"{a}}rinen(2010)]%
        {DBLP:journals/jmlr/GutmannH10}
\bibfield{author}{\bibinfo{person}{Michael Gutmann} {and} \bibinfo{person}{Aapo Hyv{\"{a}}rinen}.} \bibinfo{year}{2010}\natexlab{}.
\newblock \showarticletitle{Noise-contrastive estimation: {A} new estimation principle for unnormalized statistical models}. In \bibinfo{booktitle}{\emph{Proceedings of the Thirteenth International Conference on Artificial Intelligence and Statistics, {AISTATS} 2010, Chia Laguna Resort, Sardinia, Italy, May 13-15, 2010}} \emph{(\bibinfo{series}{{JMLR} Proceedings}, Vol.~\bibinfo{volume}{9})}, \bibfield{editor}{\bibinfo{person}{Yee~Whye Teh} {and} \bibinfo{person}{D.~Mike Titterington}} (Eds.). \bibinfo{publisher}{JMLR.org}, \bibinfo{pages}{297--304}.
\newblock
\urldef\tempurl%
\url{http://proceedings.mlr.press/v9/gutmann10a.html}
\showURL{%
\tempurl}


\bibitem[Hao et~al\mbox{.}(2023)]%
        {DBLP:journals/tkde/HaoZZLSXLZ23}
\bibfield{author}{\bibinfo{person}{Yongjing Hao}, \bibinfo{person}{Tingting Zhang}, \bibinfo{person}{Pengpeng Zhao}, \bibinfo{person}{Yanchi Liu}, \bibinfo{person}{Victor~S. Sheng}, \bibinfo{person}{Jiajie Xu}, \bibinfo{person}{Guanfeng Liu}, {and} \bibinfo{person}{Xiaofang Zhou}.} \bibinfo{year}{2023}\natexlab{}.
\newblock \showarticletitle{Feature-Level Deeper Self-Attention Network With Contrastive Learning for Sequential Recommendation}.
\newblock \bibinfo{journal}{\emph{{IEEE} Trans. Knowl. Data Eng.}} \bibinfo{volume}{35}, \bibinfo{number}{10} (\bibinfo{year}{2023}), \bibinfo{pages}{10112--10124}.
\newblock
\href{https://doi.org/10.1109/TKDE.2023.3250463}{doi:\nolinkurl{10.1109/TKDE.2023.3250463}}


\bibitem[Harte et~al\mbox{.}(2023)]%
        {DBLP:conf/recsys/HarteZLKJF23}
\bibfield{author}{\bibinfo{person}{Jesse Harte}, \bibinfo{person}{Wouter Zorgdrager}, \bibinfo{person}{Panos Louridas}, \bibinfo{person}{Asterios Katsifodimos}, \bibinfo{person}{Dietmar Jannach}, {and} \bibinfo{person}{Marios Fragkoulis}.} \bibinfo{year}{2023}\natexlab{}.
\newblock \showarticletitle{Leveraging Large Language Models for Sequential Recommendation}. In \bibinfo{booktitle}{\emph{Proceedings of the 17th {ACM} Conference on Recommender Systems, RecSys 2023, Singapore, Singapore, September 18-22, 2023}}, \bibfield{editor}{\bibinfo{person}{Jie Zhang}, \bibinfo{person}{Li~Chen}, \bibinfo{person}{Shlomo Berkovsky}, \bibinfo{person}{Min Zhang}, \bibinfo{person}{Tommaso~Di Noia}, \bibinfo{person}{Justin Basilico}, \bibinfo{person}{Luiz Pizzato}, {and} \bibinfo{person}{Yang Song}} (Eds.). \bibinfo{publisher}{{ACM}}, \bibinfo{pages}{1096--1102}.
\newblock
\href{https://doi.org/10.1145/3604915.3610639}{doi:\nolinkurl{10.1145/3604915.3610639}}


\bibitem[Hidasi et~al\mbox{.}(2016)]%
        {DBLP:journals/corr/HidasiKBT15}
\bibfield{author}{\bibinfo{person}{Bal{\'{a}}zs Hidasi}, \bibinfo{person}{Alexandros Karatzoglou}, \bibinfo{person}{Linas Baltrunas}, {and} \bibinfo{person}{Domonkos Tikk}.} \bibinfo{year}{2016}\natexlab{}.
\newblock \showarticletitle{Session-based Recommendations with Recurrent Neural Networks}. In \bibinfo{booktitle}{\emph{4th International Conference on Learning Representations, {ICLR} 2016, San Juan, Puerto Rico, May 2-4, 2016, Conference Track Proceedings}}, \bibfield{editor}{\bibinfo{person}{Yoshua Bengio} {and} \bibinfo{person}{Yann LeCun}} (Eds.).
\newblock
\urldef\tempurl%
\url{http://arxiv.org/abs/1511.06939}
\showURL{%
\tempurl}


\bibitem[Hou et~al\mbox{.}(2024)]%
        {DBLP:journals/corr/abs-2403-03952}
\bibfield{author}{\bibinfo{person}{Yupeng Hou}, \bibinfo{person}{Jiacheng Li}, \bibinfo{person}{Zhankui He}, \bibinfo{person}{An Yan}, \bibinfo{person}{Xiusi Chen}, {and} \bibinfo{person}{Julian~J. McAuley}.} \bibinfo{year}{2024}\natexlab{}.
\newblock \showarticletitle{Bridging Language and Items for Retrieval and Recommendation}.
\newblock \bibinfo{journal}{\emph{CoRR}}  \bibinfo{volume}{abs/2403.03952} (\bibinfo{year}{2024}).
\newblock
\href{https://doi.org/10.48550/ARXIV.2403.03952}{doi:\nolinkurl{10.48550/ARXIV.2403.03952}}
\showeprint[arXiv]{2403.03952}


\bibitem[Hua et~al\mbox{.}(2023)]%
        {DBLP:conf/sigir-ap/HuaXGZ23}
\bibfield{author}{\bibinfo{person}{Wenyue Hua}, \bibinfo{person}{Shuyuan Xu}, \bibinfo{person}{Yingqiang Ge}, {and} \bibinfo{person}{Yongfeng Zhang}.} \bibinfo{year}{2023}\natexlab{}.
\newblock \showarticletitle{How to Index Item IDs for Recommendation Foundation Models}. In \bibinfo{booktitle}{\emph{Annual International {ACM} {SIGIR} Conference on Research and Development in Information Retrieval in the Asia Pacific Region, {SIGIR-AP} 2023, Beijing, China, November 26-28, 2023}}, \bibfield{editor}{\bibinfo{person}{Qingyao Ai}, \bibinfo{person}{Yiqin Liu}, \bibinfo{person}{Alistair Moffat}, \bibinfo{person}{Xuanjing Huang}, \bibinfo{person}{Tetsuya Sakai}, {and} \bibinfo{person}{Justin Zobel}} (Eds.). \bibinfo{publisher}{{ACM}}, \bibinfo{pages}{195--204}.
\newblock
\href{https://doi.org/10.1145/3624918.3625339}{doi:\nolinkurl{10.1145/3624918.3625339}}


\bibitem[Kang and McAuley(2018)]%
        {sasrec}
\bibfield{author}{\bibinfo{person}{Wang{-}Cheng Kang} {and} \bibinfo{person}{Julian~J. McAuley}.} \bibinfo{year}{2018}\natexlab{}.
\newblock \showarticletitle{Self-Attentive Sequential Recommendation}. In \bibinfo{booktitle}{\emph{{IEEE} International Conference on Data Mining, {ICDM} 2018, Singapore, November 17-20, 2018}}. \bibinfo{publisher}{{IEEE} Computer Society}, \bibinfo{pages}{197--206}.
\newblock
\href{https://doi.org/10.1109/ICDM.2018.00035}{doi:\nolinkurl{10.1109/ICDM.2018.00035}}


\bibitem[Li et~al\mbox{.}(2023a)]%
        {DBLP:journals/corr/abs-2309-10706}
\bibfield{author}{\bibinfo{person}{Juntao Li}, \bibinfo{person}{Zecheng Tang}, \bibinfo{person}{Yuyang Ding}, \bibinfo{person}{Pinzheng Wang}, \bibinfo{person}{Pei Guo}, \bibinfo{person}{Wangjie You}, \bibinfo{person}{Dan Qiao}, \bibinfo{person}{Wenliang Chen}, \bibinfo{person}{Guohong Fu}, \bibinfo{person}{Qiaoming Zhu}, \bibinfo{person}{Guodong Zhou}, {and} \bibinfo{person}{Min Zhang}.} \bibinfo{year}{2023}\natexlab{a}.
\newblock \showarticletitle{OpenBA: An Open-sourced 15B Bilingual Asymmetric seq2seq Model Pre-trained from Scratch}.
\newblock \bibinfo{journal}{\emph{CoRR}}  \bibinfo{volume}{abs/2309.10706} (\bibinfo{year}{2023}).
\newblock
\href{https://doi.org/10.48550/ARXIV.2309.10706}{doi:\nolinkurl{10.48550/ARXIV.2309.10706}}
\showeprint[arXiv]{2309.10706}


\bibitem[Li et~al\mbox{.}(2023b)]%
        {gpt4rec}
\bibfield{author}{\bibinfo{person}{Jinming Li}, \bibinfo{person}{Wentao Zhang}, \bibinfo{person}{Tian Wang}, \bibinfo{person}{Guanglei Xiong}, \bibinfo{person}{Alan Lu}, {and} \bibinfo{person}{Gerard Medioni}.} \bibinfo{year}{2023}\natexlab{b}.
\newblock \showarticletitle{GPT4Rec: {A} Generative Framework for Personalized Recommendation and User Interests Interpretation}. In \bibinfo{booktitle}{\emph{Proceedings of the 2023 {SIGIR} Workshop on eCommerce co-located with the 46th International {ACM} {SIGIR} Conference on Research and Development in Information Retrieval {(SIGIR} 2023), Taipei, Taiwan, July 27, 2023}} \emph{(\bibinfo{series}{{CEUR} Workshop Proceedings}, Vol.~\bibinfo{volume}{3589})}, \bibfield{editor}{\bibinfo{person}{Surya Kallumadi}, \bibinfo{person}{Yubin Kim}, \bibinfo{person}{Tracy~Holloway King}, \bibinfo{person}{Shervin Malmasi}, \bibinfo{person}{Maarten de~Rijke}, {and} \bibinfo{person}{Jacopo Tagliabue}} (Eds.). \bibinfo{publisher}{CEUR-WS.org}.
\newblock
\urldef\tempurl%
\url{https://ceur-ws.org/Vol-3589/paper\_2.pdf}
\showURL{%
\tempurl}


\bibitem[Liu et~al\mbox{.}(2024b)]%
        {mmgrec}
\bibfield{author}{\bibinfo{person}{Han Liu}, \bibinfo{person}{Yinwei Wei}, \bibinfo{person}{Xuemeng Song}, \bibinfo{person}{Weili Guan}, \bibinfo{person}{Yuan{-}Fang Li}, {and} \bibinfo{person}{Liqiang Nie}.} \bibinfo{year}{2024}\natexlab{b}.
\newblock \showarticletitle{MMGRec: Multimodal Generative Recommendation with Transformer Model}.
\newblock \bibinfo{journal}{\emph{CoRR}}  \bibinfo{volume}{abs/2404.16555} (\bibinfo{year}{2024}).
\newblock
\href{https://doi.org/10.48550/ARXIV.2404.16555}{doi:\nolinkurl{10.48550/ARXIV.2404.16555}}
\showeprint[arXiv]{2404.16555}


\bibitem[Liu et~al\mbox{.}(2024a)]%
        {mbgr}
\bibfield{author}{\bibinfo{person}{Zihan Liu}, \bibinfo{person}{Yupeng Hou}, {and} \bibinfo{person}{Julian~J. McAuley}.} \bibinfo{year}{2024}\natexlab{a}.
\newblock \showarticletitle{Multi-Behavior Generative Recommendation}. In \bibinfo{booktitle}{\emph{Proceedings of the 33rd {ACM} International Conference on Information and Knowledge Management, {CIKM} 2024, Boise, ID, USA, October 21-25, 2024}}, \bibfield{editor}{\bibinfo{person}{Edoardo Serra} {and} \bibinfo{person}{Francesca Spezzano}} (Eds.). \bibinfo{publisher}{{ACM}}, \bibinfo{pages}{1575--1585}.
\newblock
\href{https://doi.org/10.1145/3627673.3679730}{doi:\nolinkurl{10.1145/3627673.3679730}}


\bibitem[Ma et~al\mbox{.}(2019)]%
        {DBLP:conf/kdd/MaKL19}
\bibfield{author}{\bibinfo{person}{Chen Ma}, \bibinfo{person}{Peng Kang}, {and} \bibinfo{person}{Xue Liu}.} \bibinfo{year}{2019}\natexlab{}.
\newblock \showarticletitle{Hierarchical Gating Networks for Sequential Recommendation}. In \bibinfo{booktitle}{\emph{Proceedings of the 25th {ACM} {SIGKDD} International Conference on Knowledge Discovery {\&} Data Mining, {KDD} 2019, Anchorage, AK, USA, August 4-8, 2019}}, \bibfield{editor}{\bibinfo{person}{Ankur Teredesai}, \bibinfo{person}{Vipin Kumar}, \bibinfo{person}{Ying Li}, \bibinfo{person}{R{\'{o}}mer Rosales}, \bibinfo{person}{Evimaria Terzi}, {and} \bibinfo{person}{George Karypis}} (Eds.). \bibinfo{publisher}{{ACM}}, \bibinfo{pages}{825--833}.
\newblock
\href{https://doi.org/10.1145/3292500.3330984}{doi:\nolinkurl{10.1145/3292500.3330984}}


\bibitem[Petrov and Macdonald(2023)]%
        {gptrec}
\bibfield{author}{\bibinfo{person}{Aleksandr~V. Petrov} {and} \bibinfo{person}{Craig Macdonald}.} \bibinfo{year}{2023}\natexlab{}.
\newblock \showarticletitle{Generative Sequential Recommendation with GPTRec}.
\newblock \bibinfo{journal}{\emph{CoRR}}  \bibinfo{volume}{abs/2306.11114} (\bibinfo{year}{2023}).
\newblock
\href{https://doi.org/10.48550/ARXIV.2306.11114}{doi:\nolinkurl{10.48550/ARXIV.2306.11114}}
\showeprint[arXiv]{2306.11114}


\bibitem[Qu et~al\mbox{.}(2024)]%
        {tokenrec}
\bibfield{author}{\bibinfo{person}{Haohao Qu}, \bibinfo{person}{Wenqi Fan}, \bibinfo{person}{Zihuai Zhao}, {and} \bibinfo{person}{Qing Li}.} \bibinfo{year}{2024}\natexlab{}.
\newblock \showarticletitle{TokenRec: Learning to Tokenize {ID} for LLM-based Generative Recommendation}.
\newblock \bibinfo{journal}{\emph{CoRR}}  \bibinfo{volume}{abs/2406.10450} (\bibinfo{year}{2024}).
\newblock
\href{https://doi.org/10.48550/ARXIV.2406.10450}{doi:\nolinkurl{10.48550/ARXIV.2406.10450}}
\showeprint[arXiv]{2406.10450}


\bibitem[Raffel et~al\mbox{.}(2020)]%
        {t5}
\bibfield{author}{\bibinfo{person}{Colin Raffel}, \bibinfo{person}{Noam Shazeer}, \bibinfo{person}{Adam Roberts}, \bibinfo{person}{Katherine Lee}, \bibinfo{person}{Sharan Narang}, \bibinfo{person}{Michael Matena}, \bibinfo{person}{Yanqi Zhou}, \bibinfo{person}{Wei Li}, {and} \bibinfo{person}{Peter~J. Liu}.} \bibinfo{year}{2020}\natexlab{}.
\newblock \showarticletitle{Exploring the Limits of Transfer Learning with a Unified Text-to-Text Transformer}.
\newblock \bibinfo{journal}{\emph{J. Mach. Learn. Res.}}  \bibinfo{volume}{21} (\bibinfo{year}{2020}), \bibinfo{pages}{140:1--140:67}.
\newblock
\urldef\tempurl%
\url{http://jmlr.org/papers/v21/20-074.html}
\showURL{%
\tempurl}


\bibitem[Rajput et~al\mbox{.}(2023)]%
        {tiger}
\bibfield{author}{\bibinfo{person}{Shashank Rajput}, \bibinfo{person}{Nikhil Mehta}, \bibinfo{person}{Anima Singh}, \bibinfo{person}{Raghunandan~Hulikal Keshavan}, \bibinfo{person}{Trung Vu}, \bibinfo{person}{Lukasz Heldt}, \bibinfo{person}{Lichan Hong}, \bibinfo{person}{Yi Tay}, \bibinfo{person}{Vinh~Q. Tran}, \bibinfo{person}{Jonah Samost}, \bibinfo{person}{Maciej Kula}, \bibinfo{person}{Ed~H. Chi}, {and} \bibinfo{person}{Mahesh Sathiamoorthy}.} \bibinfo{year}{2023}\natexlab{}.
\newblock \showarticletitle{Recommender Systems with Generative Retrieval}. In \bibinfo{booktitle}{\emph{Advances in Neural Information Processing Systems 36: Annual Conference on Neural Information Processing Systems 2023, NeurIPS 2023, New Orleans, LA, USA, December 10 - 16, 2023}}, \bibfield{editor}{\bibinfo{person}{Alice Oh}, \bibinfo{person}{Tristan Naumann}, \bibinfo{person}{Amir Globerson}, \bibinfo{person}{Kate Saenko}, \bibinfo{person}{Moritz Hardt}, {and} \bibinfo{person}{Sergey Levine}} (Eds.).
\newblock
\urldef\tempurl%
\url{http://papers.nips.cc/paper\_files/paper/2023/hash/20dcab0f14046a5c6b02b61da9f13229-Abstract-Conference.html}
\showURL{%
\tempurl}


\bibitem[Rendle et~al\mbox{.}(2010)]%
        {DBLP:conf/www/RendleFS10}
\bibfield{author}{\bibinfo{person}{Steffen Rendle}, \bibinfo{person}{Christoph Freudenthaler}, {and} \bibinfo{person}{Lars Schmidt{-}Thieme}.} \bibinfo{year}{2010}\natexlab{}.
\newblock \showarticletitle{Factorizing personalized Markov chains for next-basket recommendation}. In \bibinfo{booktitle}{\emph{Proceedings of the 19th International Conference on World Wide Web, {WWW} 2010, Raleigh, North Carolina, USA, April 26-30, 2010}}, \bibfield{editor}{\bibinfo{person}{Michael Rappa}, \bibinfo{person}{Paul Jones}, \bibinfo{person}{Juliana Freire}, {and} \bibinfo{person}{Soumen Chakrabarti}} (Eds.). \bibinfo{publisher}{{ACM}}, \bibinfo{pages}{811--820}.
\newblock
\href{https://doi.org/10.1145/1772690.1772773}{doi:\nolinkurl{10.1145/1772690.1772773}}


\bibitem[Si et~al\mbox{.}(2024)]%
        {seater}
\bibfield{author}{\bibinfo{person}{Zihua Si}, \bibinfo{person}{Zhongxiang Sun}, \bibinfo{person}{Jiale Chen}, \bibinfo{person}{Guozhang Chen}, \bibinfo{person}{Xiaoxue Zang}, \bibinfo{person}{Kai Zheng}, \bibinfo{person}{Yang Song}, \bibinfo{person}{Xiao Zhang}, \bibinfo{person}{Jun Xu}, {and} \bibinfo{person}{Kun Gai}.} \bibinfo{year}{2024}\natexlab{}.
\newblock \showarticletitle{Generative Retrieval with Semantic Tree-Structured Identifiers and Contrastive Learning}. In \bibinfo{booktitle}{\emph{Proceedings of the 2024 Annual International ACM SIGIR Conference on Research and Development in Information Retrieval in the Asia Pacific Region}} (Tokyo, Japan) \emph{(\bibinfo{series}{SIGIR-AP 2024})}. \bibinfo{publisher}{Association for Computing Machinery}, \bibinfo{address}{New York, NY, USA}, \bibinfo{pages}{154–163}.
\newblock
\showISBNx{9798400707247}
\href{https://doi.org/10.1145/3673791.3698408}{doi:\nolinkurl{10.1145/3673791.3698408}}


\bibitem[Sun et~al\mbox{.}(2019)]%
        {bert4rec}
\bibfield{author}{\bibinfo{person}{Fei Sun}, \bibinfo{person}{Jun Liu}, \bibinfo{person}{Jian Wu}, \bibinfo{person}{Changhua Pei}, \bibinfo{person}{Xiao Lin}, \bibinfo{person}{Wenwu Ou}, {and} \bibinfo{person}{Peng Jiang}.} \bibinfo{year}{2019}\natexlab{}.
\newblock \showarticletitle{BERT4Rec: Sequential Recommendation with Bidirectional Encoder Representations from Transformer}. In \bibinfo{booktitle}{\emph{Proceedings of the 28th {ACM} International Conference on Information and Knowledge Management, {CIKM} 2019, Beijing, China, November 3-7, 2019}}, \bibfield{editor}{\bibinfo{person}{Wenwu Zhu}, \bibinfo{person}{Dacheng Tao}, \bibinfo{person}{Xueqi Cheng}, \bibinfo{person}{Peng Cui}, \bibinfo{person}{Elke~A. Rundensteiner}, \bibinfo{person}{David Carmel}, \bibinfo{person}{Qi~He}, {and} \bibinfo{person}{Jeffrey~Xu Yu}} (Eds.). \bibinfo{publisher}{{ACM}}, \bibinfo{pages}{1441--1450}.
\newblock
\href{https://doi.org/10.1145/3357384.3357895}{doi:\nolinkurl{10.1145/3357384.3357895}}


\bibitem[Sun et~al\mbox{.}(2023)]%
        {DBLP:conf/nips/0001YCWZRCYRR23}
\bibfield{author}{\bibinfo{person}{Weiwei Sun}, \bibinfo{person}{Lingyong Yan}, \bibinfo{person}{Zheng Chen}, \bibinfo{person}{Shuaiqiang Wang}, \bibinfo{person}{Haichao Zhu}, \bibinfo{person}{Pengjie Ren}, \bibinfo{person}{Zhumin Chen}, \bibinfo{person}{Dawei Yin}, \bibinfo{person}{Maarten de Rijke}, {and} \bibinfo{person}{Zhaochun Ren}.} \bibinfo{year}{2023}\natexlab{}.
\newblock \showarticletitle{Learning to Tokenize for Generative Retrieval}. In \bibinfo{booktitle}{\emph{Advances in Neural Information Processing Systems 36: Annual Conference on Neural Information Processing Systems 2023, NeurIPS 2023, New Orleans, LA, USA, December 10 - 16, 2023}}, \bibfield{editor}{\bibinfo{person}{Alice Oh}, \bibinfo{person}{Tristan Naumann}, \bibinfo{person}{Amir Globerson}, \bibinfo{person}{Kate Saenko}, \bibinfo{person}{Moritz Hardt}, {and} \bibinfo{person}{Sergey Levine}} (Eds.).
\newblock
\urldef\tempurl%
\url{http://papers.nips.cc/paper\_files/paper/2023/hash/91228b942a4528cdae031c1b68b127e8-Abstract-Conference.html}
\showURL{%
\tempurl}


\bibitem[Tan et~al\mbox{.}(2024)]%
        {idgenrec}
\bibfield{author}{\bibinfo{person}{Juntao Tan}, \bibinfo{person}{Shuyuan Xu}, \bibinfo{person}{Wenyue Hua}, \bibinfo{person}{Yingqiang Ge}, \bibinfo{person}{Zelong Li}, {and} \bibinfo{person}{Yongfeng Zhang}.} \bibinfo{year}{2024}\natexlab{}.
\newblock \showarticletitle{IDGenRec: LLM-RecSys Alignment with Textual {ID} Learning}. In \bibinfo{booktitle}{\emph{Proceedings of the 47th International {ACM} {SIGIR} Conference on Research and Development in Information Retrieval, {SIGIR} 2024, Washington DC, USA, July 14-18, 2024}}, \bibfield{editor}{\bibinfo{person}{Grace~Hui Yang}, \bibinfo{person}{Hongning Wang}, \bibinfo{person}{Sam Han}, \bibinfo{person}{Claudia Hauff}, \bibinfo{person}{Guido Zuccon}, {and} \bibinfo{person}{Yi~Zhang}} (Eds.). \bibinfo{publisher}{{ACM}}, \bibinfo{pages}{355--364}.
\newblock
\href{https://doi.org/10.1145/3626772.3657821}{doi:\nolinkurl{10.1145/3626772.3657821}}


\bibitem[Tan et~al\mbox{.}(2016)]%
        {DBLP:conf/recsys/TanXL16}
\bibfield{author}{\bibinfo{person}{Yong~Kiam Tan}, \bibinfo{person}{Xinxing Xu}, {and} \bibinfo{person}{Yong Liu}.} \bibinfo{year}{2016}\natexlab{}.
\newblock \showarticletitle{Improved Recurrent Neural Networks for Session-based Recommendations}. In \bibinfo{booktitle}{\emph{Proceedings of the 1st Workshop on Deep Learning for Recommender Systems, DLRS@RecSys 2016, Boston, MA, USA, September 15, 2016}}, \bibfield{editor}{\bibinfo{person}{Alexandros Karatzoglou}, \bibinfo{person}{Bal{\'{a}}zs Hidasi}, \bibinfo{person}{Domonkos Tikk}, \bibinfo{person}{Oren~Sar Shalom}, \bibinfo{person}{Haggai Roitman}, \bibinfo{person}{Bracha Shapira}, {and} \bibinfo{person}{Lior Rokach}} (Eds.). \bibinfo{publisher}{{ACM}}, \bibinfo{pages}{17--22}.
\newblock
\href{https://doi.org/10.1145/2988450.2988452}{doi:\nolinkurl{10.1145/2988450.2988452}}


\bibitem[Tang and Wang(2018)]%
        {DBLP:conf/wsdm/TangW18}
\bibfield{author}{\bibinfo{person}{Jiaxi Tang} {and} \bibinfo{person}{Ke Wang}.} \bibinfo{year}{2018}\natexlab{}.
\newblock \showarticletitle{Personalized Top-N Sequential Recommendation via Convolutional Sequence Embedding}. In \bibinfo{booktitle}{\emph{Proceedings of the Eleventh {ACM} International Conference on Web Search and Data Mining, {WSDM} 2018, Marina Del Rey, CA, USA, February 5-9, 2018}}, \bibfield{editor}{\bibinfo{person}{Yi~Chang}, \bibinfo{person}{Chengxiang Zhai}, \bibinfo{person}{Yan Liu}, {and} \bibinfo{person}{Yoelle Maarek}} (Eds.). \bibinfo{publisher}{{ACM}}, \bibinfo{pages}{565--573}.
\newblock
\href{https://doi.org/10.1145/3159652.3159656}{doi:\nolinkurl{10.1145/3159652.3159656}}


\bibitem[van~den Oord et~al\mbox{.}(2017)]%
        {DBLP:conf/nips/OordVK17}
\bibfield{author}{\bibinfo{person}{A{\"{a}}ron van~den Oord}, \bibinfo{person}{Oriol Vinyals}, {and} \bibinfo{person}{Koray Kavukcuoglu}.} \bibinfo{year}{2017}\natexlab{}.
\newblock \showarticletitle{Neural Discrete Representation Learning}. In \bibinfo{booktitle}{\emph{Advances in Neural Information Processing Systems 30: Annual Conference on Neural Information Processing Systems 2017, December 4-9, 2017, Long Beach, CA, {USA}}}, \bibfield{editor}{\bibinfo{person}{Isabelle Guyon}, \bibinfo{person}{Ulrike von Luxburg}, \bibinfo{person}{Samy Bengio}, \bibinfo{person}{Hanna~M. Wallach}, \bibinfo{person}{Rob Fergus}, \bibinfo{person}{S.~V.~N. Vishwanathan}, {and} \bibinfo{person}{Roman Garnett}} (Eds.). \bibinfo{pages}{6306--6315}.
\newblock
\urldef\tempurl%
\url{https://proceedings.neurips.cc/paper/2017/hash/7a98af17e63a0ac09ce2e96d03992fbc-Abstract.html}
\showURL{%
\tempurl}


\bibitem[van~der Maaten and Hinton(2008)]%
        {tsne}
\bibfield{author}{\bibinfo{person}{Laurens van~der Maaten} {and} \bibinfo{person}{Geoffrey Hinton}.} \bibinfo{year}{2008}\natexlab{}.
\newblock \showarticletitle{Visualizing Data using t-SNE}.
\newblock \bibinfo{journal}{\emph{Journal of Machine Learning Research}} \bibinfo{volume}{9}, \bibinfo{number}{86} (\bibinfo{year}{2008}), \bibinfo{pages}{2579--2605}.
\newblock
\urldef\tempurl%
\url{http://jmlr.org/papers/v9/vandermaaten08a.html}
\showURL{%
\tempurl}


\bibitem[Vaswani et~al\mbox{.}(2017)]%
        {transformer}
\bibfield{author}{\bibinfo{person}{Ashish Vaswani}, \bibinfo{person}{Noam Shazeer}, \bibinfo{person}{Niki Parmar}, \bibinfo{person}{Jakob Uszkoreit}, \bibinfo{person}{Llion Jones}, \bibinfo{person}{Aidan~N. Gomez}, \bibinfo{person}{Lukasz Kaiser}, {and} \bibinfo{person}{Illia Polosukhin}.} \bibinfo{year}{2017}\natexlab{}.
\newblock \showarticletitle{Attention is All you Need}. In \bibinfo{booktitle}{\emph{Advances in Neural Information Processing Systems 30: Annual Conference on Neural Information Processing Systems 2017, December 4-9, 2017, Long Beach, CA, {USA}}}, \bibfield{editor}{\bibinfo{person}{Isabelle Guyon}, \bibinfo{person}{Ulrike von Luxburg}, \bibinfo{person}{Samy Bengio}, \bibinfo{person}{Hanna~M. Wallach}, \bibinfo{person}{Rob Fergus}, \bibinfo{person}{S.~V.~N. Vishwanathan}, {and} \bibinfo{person}{Roman Garnett}} (Eds.). \bibinfo{pages}{5998--6008}.
\newblock
\urldef\tempurl%
\url{https://proceedings.neurips.cc/paper/2017/hash/3f5ee243547dee91fbd053c1c4a845aa-Abstract.html}
\showURL{%
\tempurl}


\bibitem[Wang et~al\mbox{.}(2024a)]%
        {letter}
\bibfield{author}{\bibinfo{person}{Wenjie Wang}, \bibinfo{person}{Honghui Bao}, \bibinfo{person}{Xinyu Lin}, \bibinfo{person}{Jizhi Zhang}, \bibinfo{person}{Yongqi Li}, \bibinfo{person}{Fuli Feng}, \bibinfo{person}{See{-}Kiong Ng}, {and} \bibinfo{person}{Tat{-}Seng Chua}.} \bibinfo{year}{2024}\natexlab{a}.
\newblock \showarticletitle{Learnable Item Tokenization for Generative Recommendation}. In \bibinfo{booktitle}{\emph{Proceedings of the 33rd {ACM} International Conference on Information and Knowledge Management, {CIKM} 2024, Boise, ID, USA, October 21-25, 2024}}, \bibfield{editor}{\bibinfo{person}{Edoardo Serra} {and} \bibinfo{person}{Francesca Spezzano}} (Eds.). \bibinfo{publisher}{{ACM}}, \bibinfo{pages}{2400--2409}.
\newblock
\href{https://doi.org/10.1145/3627673.3679569}{doi:\nolinkurl{10.1145/3627673.3679569}}


\bibitem[Wang et~al\mbox{.}(2024b)]%
        {DBLP:journals/corr/abs-2403-18480}
\bibfield{author}{\bibinfo{person}{Yidan Wang}, \bibinfo{person}{Zhaochun Ren}, \bibinfo{person}{Weiwei Sun}, \bibinfo{person}{Jiyuan Yang}, \bibinfo{person}{Zhixiang Liang}, \bibinfo{person}{Xin Chen}, \bibinfo{person}{Ruobing Xie}, \bibinfo{person}{Su Yan}, \bibinfo{person}{Xu Zhang}, \bibinfo{person}{Pengjie Ren}, \bibinfo{person}{Zhumin Chen}, {and} \bibinfo{person}{Xin Xin}.} \bibinfo{year}{2024}\natexlab{b}.
\newblock \showarticletitle{Enhanced Generative Recommendation via Content and Collaboration Integration}.
\newblock \bibinfo{journal}{\emph{CoRR}}  \bibinfo{volume}{abs/2403.18480} (\bibinfo{year}{2024}).
\newblock
\href{https://doi.org/10.48550/ARXIV.2403.18480}{doi:\nolinkurl{10.48550/ARXIV.2403.18480}}
\showeprint[arXiv]{2403.18480}


\bibitem[Wang et~al\mbox{.}(2024c)]%
        {eager}
\bibfield{author}{\bibinfo{person}{Ye Wang}, \bibinfo{person}{Jiahao Xun}, \bibinfo{person}{Minjie Hong}, \bibinfo{person}{Jieming Zhu}, \bibinfo{person}{Tao Jin}, \bibinfo{person}{Wang Lin}, \bibinfo{person}{Haoyuan Li}, \bibinfo{person}{Linjun Li}, \bibinfo{person}{Yan Xia}, \bibinfo{person}{Zhou Zhao}, {and} \bibinfo{person}{Zhenhua Dong}.} \bibinfo{year}{2024}\natexlab{c}.
\newblock \showarticletitle{{EAGER:} Two-Stream Generative Recommender with Behavior-Semantic Collaboration}. In \bibinfo{booktitle}{\emph{Proceedings of the 30th {ACM} {SIGKDD} Conference on Knowledge Discovery and Data Mining, {KDD} 2024, Barcelona, Spain, August 25-29, 2024}}, \bibfield{editor}{\bibinfo{person}{Ricardo Baeza{-}Yates} {and} \bibinfo{person}{Francesco Bonchi}} (Eds.). \bibinfo{publisher}{{ACM}}, \bibinfo{pages}{3245--3254}.
\newblock
\href{https://doi.org/10.1145/3637528.3671775}{doi:\nolinkurl{10.1145/3637528.3671775}}


\bibitem[Wu et~al\mbox{.}(2019)]%
        {DBLP:conf/aaai/WuT0WXT19}
\bibfield{author}{\bibinfo{person}{Shu Wu}, \bibinfo{person}{Yuyuan Tang}, \bibinfo{person}{Yanqiao Zhu}, \bibinfo{person}{Liang Wang}, \bibinfo{person}{Xing Xie}, {and} \bibinfo{person}{Tieniu Tan}.} \bibinfo{year}{2019}\natexlab{}.
\newblock \showarticletitle{Session-Based Recommendation with Graph Neural Networks}. In \bibinfo{booktitle}{\emph{The Thirty-Third {AAAI} Conference on Artificial Intelligence, {AAAI} 2019, The Thirty-First Innovative Applications of Artificial Intelligence Conference, {IAAI} 2019, The Ninth {AAAI} Symposium on Educational Advances in Artificial Intelligence, {EAAI} 2019, Honolulu, Hawaii, USA, January 27 - February 1, 2019}}. \bibinfo{publisher}{{AAAI} Press}, \bibinfo{pages}{346--353}.
\newblock
\href{https://doi.org/10.1609/AAAI.V33I01.3301346}{doi:\nolinkurl{10.1609/AAAI.V33I01.3301346}}


\bibitem[Xie et~al\mbox{.}(2022)]%
        {DBLP:conf/sigir/XieZK22}
\bibfield{author}{\bibinfo{person}{Yueqi Xie}, \bibinfo{person}{Peilin Zhou}, {and} \bibinfo{person}{Sunghun Kim}.} \bibinfo{year}{2022}\natexlab{}.
\newblock \showarticletitle{Decoupled Side Information Fusion for Sequential Recommendation}. In \bibinfo{booktitle}{\emph{{SIGIR} '22: The 45th International {ACM} {SIGIR} Conference on Research and Development in Information Retrieval, Madrid, Spain, July 11 - 15, 2022}}, \bibfield{editor}{\bibinfo{person}{Enrique Amig{\'{o}}}, \bibinfo{person}{Pablo Castells}, \bibinfo{person}{Julio Gonzalo}, \bibinfo{person}{Ben Carterette}, \bibinfo{person}{J.~Shane Culpepper}, {and} \bibinfo{person}{Gabriella Kazai}} (Eds.). \bibinfo{publisher}{{ACM}}, \bibinfo{pages}{1611--1621}.
\newblock
\href{https://doi.org/10.1145/3477495.3531963}{doi:\nolinkurl{10.1145/3477495.3531963}}


\bibitem[Yang et~al\mbox{.}(2023)]%
        {autoindexer}
\bibfield{author}{\bibinfo{person}{Tianchi Yang}, \bibinfo{person}{Minghui Song}, \bibinfo{person}{Zihan Zhang}, \bibinfo{person}{Haizhen Huang}, \bibinfo{person}{Weiwei Deng}, \bibinfo{person}{Feng Sun}, {and} \bibinfo{person}{Qi Zhang}.} \bibinfo{year}{2023}\natexlab{}.
\newblock \showarticletitle{Auto Search Indexer for End-to-End Document Retrieval}. In \bibinfo{booktitle}{\emph{Findings of the Association for Computational Linguistics: {EMNLP} 2023, Singapore, December 6-10, 2023}}, \bibfield{editor}{\bibinfo{person}{Houda Bouamor}, \bibinfo{person}{Juan Pino}, {and} \bibinfo{person}{Kalika Bali}} (Eds.). \bibinfo{publisher}{Association for Computational Linguistics}, \bibinfo{pages}{6955--6970}.
\newblock
\href{https://doi.org/10.18653/V1/2023.FINDINGS-EMNLP.464}{doi:\nolinkurl{10.18653/V1/2023.FINDINGS-EMNLP.464}}


\bibitem[Yue et~al\mbox{.}(2023)]%
        {DBLP:journals/corr/abs-2311-02089}
\bibfield{author}{\bibinfo{person}{Zhenrui Yue}, \bibinfo{person}{Sara Rabhi}, \bibinfo{person}{Gabriel de Souza Pereira~Moreira}, \bibinfo{person}{Dong Wang}, {and} \bibinfo{person}{Even Oldridge}.} \bibinfo{year}{2023}\natexlab{}.
\newblock \showarticletitle{LlamaRec: Two-Stage Recommendation using Large Language Models for Ranking}.
\newblock \bibinfo{journal}{\emph{CoRR}}  \bibinfo{volume}{abs/2311.02089} (\bibinfo{year}{2023}).
\newblock
\href{https://doi.org/10.48550/ARXIV.2311.02089}{doi:\nolinkurl{10.48550/ARXIV.2311.02089}}
\showeprint[arXiv]{2311.02089}


\bibitem[Zeghidour et~al\mbox{.}(2022)]%
        {DBLP:journals/taslp/ZeghidourLOST22}
\bibfield{author}{\bibinfo{person}{Neil Zeghidour}, \bibinfo{person}{Alejandro Luebs}, \bibinfo{person}{Ahmed Omran}, \bibinfo{person}{Jan Skoglund}, {and} \bibinfo{person}{Marco Tagliasacchi}.} \bibinfo{year}{2022}\natexlab{}.
\newblock \showarticletitle{SoundStream: An End-to-End Neural Audio Codec}.
\newblock \bibinfo{journal}{\emph{{IEEE} {ACM} Trans. Audio Speech Lang. Process.}}  \bibinfo{volume}{30} (\bibinfo{year}{2022}), \bibinfo{pages}{495--507}.
\newblock
\href{https://doi.org/10.1109/TASLP.2021.3129994}{doi:\nolinkurl{10.1109/TASLP.2021.3129994}}


\bibitem[Zhang et~al\mbox{.}(2019)]%
        {fdsa}
\bibfield{author}{\bibinfo{person}{Tingting Zhang}, \bibinfo{person}{Pengpeng Zhao}, \bibinfo{person}{Yanchi Liu}, \bibinfo{person}{Victor~S. Sheng}, \bibinfo{person}{Jiajie Xu}, \bibinfo{person}{Deqing Wang}, \bibinfo{person}{Guanfeng Liu}, {and} \bibinfo{person}{Xiaofang Zhou}.} \bibinfo{year}{2019}\natexlab{}.
\newblock \showarticletitle{Feature-level Deeper Self-Attention Network for Sequential Recommendation}. In \bibinfo{booktitle}{\emph{Proceedings of the Twenty-Eighth International Joint Conference on Artificial Intelligence, {IJCAI} 2019, Macao, China, August 10-16, 2019}}, \bibfield{editor}{\bibinfo{person}{Sarit Kraus}} (Ed.). \bibinfo{publisher}{ijcai.org}, \bibinfo{pages}{4320--4326}.
\newblock
\href{https://doi.org/10.24963/IJCAI.2019/600}{doi:\nolinkurl{10.24963/IJCAI.2019/600}}


\bibitem[Zhao et~al\mbox{.}(2022)]%
        {recbole2.0}
\bibfield{author}{\bibinfo{person}{Wayne~Xin Zhao}, \bibinfo{person}{Yupeng Hou}, \bibinfo{person}{Xingyu Pan}, \bibinfo{person}{Chen Yang}, \bibinfo{person}{Zeyu Zhang}, \bibinfo{person}{Zihan Lin}, \bibinfo{person}{Jingsen Zhang}, \bibinfo{person}{Shuqing Bian}, \bibinfo{person}{Jiakai Tang}, \bibinfo{person}{Wenqi Sun}, \bibinfo{person}{Yushuo Chen}, \bibinfo{person}{Lanling Xu}, \bibinfo{person}{Gaowei Zhang}, \bibinfo{person}{Zhen Tian}, \bibinfo{person}{Changxin Tian}, \bibinfo{person}{Shanlei Mu}, \bibinfo{person}{Xinyan Fan}, \bibinfo{person}{Xu Chen}, {and} \bibinfo{person}{Ji{-}Rong Wen}.} \bibinfo{year}{2022}\natexlab{}.
\newblock \showarticletitle{RecBole 2.0: Towards a More Up-to-Date Recommendation Library}. In \bibinfo{booktitle}{\emph{Proceedings of the 31st {ACM} International Conference on Information {\&} Knowledge Management, Atlanta, GA, USA, October 17-21, 2022}}, \bibfield{editor}{\bibinfo{person}{Mohammad~Al Hasan} {and} \bibinfo{person}{Li~Xiong}} (Eds.). \bibinfo{publisher}{{ACM}}, \bibinfo{pages}{4722--4726}.
\newblock
\href{https://doi.org/10.1145/3511808.3557680}{doi:\nolinkurl{10.1145/3511808.3557680}}


\bibitem[Zhao et~al\mbox{.}(2021)]%
        {recbole1.0}
\bibfield{author}{\bibinfo{person}{Wayne~Xin Zhao}, \bibinfo{person}{Shanlei Mu}, \bibinfo{person}{Yupeng Hou}, \bibinfo{person}{Zihan Lin}, \bibinfo{person}{Yushuo Chen}, \bibinfo{person}{Xingyu Pan}, \bibinfo{person}{Kaiyuan Li}, \bibinfo{person}{Yujie Lu}, \bibinfo{person}{Hui Wang}, \bibinfo{person}{Changxin Tian}, \bibinfo{person}{Yingqian Min}, \bibinfo{person}{Zhichao Feng}, \bibinfo{person}{Xinyan Fan}, \bibinfo{person}{Xu Chen}, \bibinfo{person}{Pengfei Wang}, \bibinfo{person}{Wendi Ji}, \bibinfo{person}{Yaliang Li}, \bibinfo{person}{Xiaoling Wang}, {and} \bibinfo{person}{Ji{-}Rong Wen}.} \bibinfo{year}{2021}\natexlab{}.
\newblock \showarticletitle{RecBole: Towards a Unified, Comprehensive and Efficient Framework for Recommendation Algorithms}. In \bibinfo{booktitle}{\emph{{CIKM} '21: The 30th {ACM} International Conference on Information and Knowledge Management, Virtual Event, Queensland, Australia, November 1 - 5, 2021}}, \bibfield{editor}{\bibinfo{person}{Gianluca Demartini}, \bibinfo{person}{Guido Zuccon}, \bibinfo{person}{J.~Shane Culpepper}, \bibinfo{person}{Zi~Huang}, {and} \bibinfo{person}{Hanghang Tong}} (Eds.). \bibinfo{publisher}{{ACM}}, \bibinfo{pages}{4653--4664}.
\newblock
\href{https://doi.org/10.1145/3459637.3482016}{doi:\nolinkurl{10.1145/3459637.3482016}}


\bibitem[Zheng et~al\mbox{.}(2024)]%
        {lcrec}
\bibfield{author}{\bibinfo{person}{Bowen Zheng}, \bibinfo{person}{Yupeng Hou}, \bibinfo{person}{Hongyu Lu}, \bibinfo{person}{Yu Chen}, \bibinfo{person}{Wayne~Xin Zhao}, \bibinfo{person}{Ming Chen}, {and} \bibinfo{person}{Ji{-}Rong Wen}.} \bibinfo{year}{2024}\natexlab{}.
\newblock \showarticletitle{Adapting Large Language Models by Integrating Collaborative Semantics for Recommendation}. In \bibinfo{booktitle}{\emph{40th {IEEE} International Conference on Data Engineering, {ICDE} 2024, Utrecht, The Netherlands, May 13-16, 2024}}. \bibinfo{publisher}{{IEEE}}, \bibinfo{pages}{1435--1448}.
\newblock
\href{https://doi.org/10.1109/ICDE60146.2024.00118}{doi:\nolinkurl{10.1109/ICDE60146.2024.00118}}


\bibitem[Zhou et~al\mbox{.}(2020)]%
        {s3rec}
\bibfield{author}{\bibinfo{person}{Kun Zhou}, \bibinfo{person}{Hui Wang}, \bibinfo{person}{Wayne~Xin Zhao}, \bibinfo{person}{Yutao Zhu}, \bibinfo{person}{Sirui Wang}, \bibinfo{person}{Fuzheng Zhang}, \bibinfo{person}{Zhongyuan Wang}, {and} \bibinfo{person}{Ji{-}Rong Wen}.} \bibinfo{year}{2020}\natexlab{}.
\newblock \showarticletitle{S3-Rec: Self-Supervised Learning for Sequential Recommendation with Mutual Information Maximization}. In \bibinfo{booktitle}{\emph{{CIKM} '20: The 29th {ACM} International Conference on Information and Knowledge Management, Virtual Event, Ireland, October 19-23, 2020}}, \bibfield{editor}{\bibinfo{person}{Mathieu d'Aquin}, \bibinfo{person}{Stefan Dietze}, \bibinfo{person}{Claudia Hauff}, \bibinfo{person}{Edward Curry}, {and} \bibinfo{person}{Philippe Cudr{\'{e}}{-}Mauroux}} (Eds.). \bibinfo{publisher}{{ACM}}, \bibinfo{pages}{1893--1902}.
\newblock
\href{https://doi.org/10.1145/3340531.3411954}{doi:\nolinkurl{10.1145/3340531.3411954}}


\bibitem[Zhou et~al\mbox{.}(2022)]%
        {DBLP:conf/www/ZhouYZW22}
\bibfield{author}{\bibinfo{person}{Kun Zhou}, \bibinfo{person}{Hui Yu}, \bibinfo{person}{Wayne~Xin Zhao}, {and} \bibinfo{person}{Ji{-}Rong Wen}.} \bibinfo{year}{2022}\natexlab{}.
\newblock \showarticletitle{Filter-enhanced {MLP} is All You Need for Sequential Recommendation}. In \bibinfo{booktitle}{\emph{{WWW} '22: The {ACM} Web Conference 2022, Virtual Event, Lyon, France, April 25 - 29, 2022}}, \bibfield{editor}{\bibinfo{person}{Fr{\'{e}}d{\'{e}}rique Laforest}, \bibinfo{person}{Rapha{\"{e}}l Troncy}, \bibinfo{person}{Elena Simperl}, \bibinfo{person}{Deepak Agarwal}, \bibinfo{person}{Aristides Gionis}, \bibinfo{person}{Ivan Herman}, {and} \bibinfo{person}{Lionel M{\'{e}}dini}} (Eds.). \bibinfo{publisher}{{ACM}}, \bibinfo{pages}{2388--2399}.
\newblock
\href{https://doi.org/10.1145/3485447.3512111}{doi:\nolinkurl{10.1145/3485447.3512111}}


\end{thebibliography}


\end{document}